\documentclass[twocolumn,apjfonts]{aastex631}

\usepackage{apjfonts}

\usepackage{graphicx}
\usepackage[suffix=]{epstopdf}
\usepackage{natbib}
\usepackage{amsmath}
\usepackage{url}
\usepackage{xspace}
\usepackage{multirow}
\usepackage{booktabs}
\usepackage{color}
\usepackage{tcolorbox}
\usepackage{placeins}

\usepackage{hyperref}

\usepackage{float}

\providecommand{\e}[1]{\ensuremath{\times 10^{#1}}}

\newcommand{\pc}{{\rm pc}\xspace}
\newcommand{\Myr}{{\rm Myr}\xspace}

\makeatletter
\newcommand{\unit}[1]{%
    \,\mathrm{#1}\checknextarg}
\newcommand{\checknextarg}{\@ifnextchar\bgroup{\gobblenextarg}{}}
\newcommand{\gobblenextarg}[1]{\,\mathrm{#1}\@ifnextchar\bgroup{\gobblenextarg}{}}
\makeatother

\newif\ifstartedinmathmode
\newcommand{\msun}{%
  \relax\ifmmode\startedinmathmodetrue\else\startedinmathmodefalse\fi
  {\ifstartedinmathmode\unit{M_{\odot}}\else$\unit{M_{\odot}}$\fi}\xspace%
}

\newif\ifstartedinmathmode
\newcommand{\Msun}{%
  \relax\ifmmode\startedinmathmodetrue\else\startedinmathmodefalse\fi
  {\ifstartedinmathmode\unit{M_{\odot}}\else$\unit{M_{\odot}}$\fi}\xspace%
}

\shorttitle{Embedded Timescales in STARFORGE}
\shortauthors{Wainer et al.}

\begin{document}

\title{The Timescales of Embedded Star Formation as Observed in STARFORGE}

\correspondingauthor{Tobin M. Wainer}
\email{tobinw@uw.edu}

\author[0000-0001-6320-2230]{Tobin M. Wainer}
\affiliation{Department of Astronomy, University of Washington, Box 351580, Seattle, WA 98195, USA}
\affiliation{Center for Computational Astrophysics, Flatiron Institute, 162 Fifth Avenue, New York, NY 10010, USA}

\author[0000-0002-1264-2006]{Julianne J.\ Dalcanton}
\affiliation{Center for Computational Astrophysics, Flatiron Institute, 162 Fifth Avenue, New York, NY 10010, USA}
\affiliation{Department of Astronomy, University of Washington, Box 351580, Seattle, WA 98195, USA}

\author[0000-0002-1655-5604]{Michael Y. Grudi\'c}
\affiliation{Center for Computational Astrophysics, Flatiron Institute,
162 Fifth Avenue, New York, NY 10010, USA}

\author[0000-0003-1252-9916]{Stella S. R. Offner}
\affiliation{Department of Astronomy, University of Texas at Austin \\
Austin, TX, 78712 USA}

\author[0000-0003-2599-7524]{Adam Smercina}\thanks{NHFP Hubble Fellow}
\affiliation{Space Telescope Science Institute, 3700 San Martin Dr., Baltimore, MD 21218, USA}

\author[0000-0002-7502-0597]{Benjamin F. Williams}
\affiliation{Department of Astronomy, University of Washington, Box 351580, Seattle, WA 98195, USA}

\author[0000-0001-6421-0953]{L. Clifton Johnson}
\affiliation{Center for Interdisciplinary Exploration and Research in Astrophysics (CIERA) and Department of Physics and Astronomy, Northwestern University, 1800 Sherman Ave., Evanston, IL 60201, USA}

\author[0000-0002-5937-9778]{J. Peltonen}
\affiliation{Department of Physics, University of Alberta, Edmonton, AB T6G 2E1, Canada}

\author[0000-0001-9605-780X]{Eric W. Koch}
\affiliation{National Radio Astronomy Observatory, 800 Bradbury SE, Suite 235, Albuquerque, NM 87106 USA}

\author[0000-0003-3205-4460]{Kartik R. Neralwar}
\affiliation{Max-Planck-Institut f\"ur Radioastronomie, Auf dem H\"ugel 69, 53121 Bonn, Germany}

\begin{abstract}
    Star formation occurs within dusty molecular clouds that are then disrupted by stellar feedback. However, the timing and physical mechanisms that govern the transition from deeply embedded to exposed stars remain uncertain. Using the STARFORGE simulations, we analyze the evolution of ``embeddedness'', identifying what drives emergence. We find the transition from embedded to exposed is fast for individual stars, within 1.3 Myr after the star reaches its maximum mass. This rapid transition is dominated by massive stars, which accrete while remaining highly obscured until their feedback eventually balances, then overcomes, the local accretion. For these massive stars, their maximum mass is reached simultaneously with their emergence. Once these stars are revealed, their localized, pre-supernova feedback then impacts the cloud, driving gas clearance. Because massive stars dominate the luminosity, their fast, local evolution dominates the light emergence from the dust. We calculate the dependence of these processes on the mass of the cloud and find that emergence always depends on when massive stars form, which scales with the cloud's free-fall time. We also measure the evolution of dust and H$\alpha$ luminosities, where for $\sim$2 Myr, these tracers outshine the emerging stellar continuum, reaching their peak when gas and dust remain tightly coupled to the massive stars. These results closely resemble observationally observed lifetimes, tying the observable dust and line emission directly to the same localized processes that drive stellar emergence, evidence that our simulated de-embedding physics is representative of real star-forming regions. 
    Thus, because the initial embedding of the most luminous stars is highly local, the emergence of stars is a faster, earlier, more local event than the overall disruption of the cloud by gas expulsion. 
\end{abstract}


\section{Introduction}

It is widely accepted that stars are born during the gravitational collapse of turbulent molecular clouds \citep[e.g.,][for review]{lada_embedded_2003}. These stars eventually emerge from the cloud, through a variety of feedback processes that clear the surrounding gas, leaving a largely un-obscured view of the newly-formed stars \citep{krause_physics_2020, pineda_bubbles_2023}.

Early on, however, accreting protostars remain enshrouded in their dusty natal clouds. These young stars are typically shielded by heavy layers of dust, making them largely invisible at optical wavelengths. At longer wavelengths however, they can be detected -- either directly as individual stars, or, indirectly through their heating of the surrounding dust. In this early phase, these collections of enshrouded stars and protostars are known as embedded clusters and are some of the earliest windows onto to the star formation process \citep{ascenso_embedded_2018}. 

This embedded stage concludes when gas is expelled from the region, exposing the stars, and consequently destroying the GMC \citep[See reviews][and references within]{krumholz_big_2014, chevance_molecular_2020, schinnerer_molecular_2024}. The primary mechanisms for this gas removal are through accretion-driven bipolar jets and outflows, radiation, and stellar winds from protostars \citep{grudic_dynamics_2022, guszejnov_effects_2022}.

Two primary parameters are thought to determine the evolution of an emerging embedded cluster: the star formation efficiency and the timescales of gas dispersal \citep{lada_embedded_2003}. Specifically, gas removal, and the timescales on which it occurs, is critical to star formation models. Once material is removed from the system, star formation is quenched, not only regulating star formation locally, but on a galactic scale \citep[e.g.,][]{krumholz_formation_2005, krumholz_star_2019}. 

Despite its importance, the underlying mechanism which govern how, and when, gas is cleared -- and the associated observability of the embedded phase -- remains poorly constrained. There remain ongoing conversations about the timescales over which this gas removal takes place, how long the newly formed stars remain embedded observationally, and what physical processes drive the clearance and its timescales \citep[e.g.,][]{mcleod_impact_2021, chevance_pre-supernova_2022, mcquaid_timescales_2024}. 

Efforts to constrain embedded cluster lifetimes have historically relied on associating young stellar clusters with their parent molecular clouds. Early studies found that clusters younger than $\sim5$ Myr were often still embedded, while older ones were not, suggesting gas is typically cleared on $\sim5$ Myr timescales \citep[e.g.,][]{leisawitz_co_1989, whitmore_what_2002,lada_embedded_2003}. More recent work extended this method to external galaxies using HST to detect young clusters and using ALMA observations to trace molecular gas or H$\alpha$ to trace ionized gas \citep[e.g.,][]{whitmore_using_2011, grasha_connecting_2018, hannon_h_2022, whitmore_empirical_2025}. These larger samples often yield shorter inferred timescales of 1–2 \Myr across diverse galactic environments.

However, a central limitation is that optical and UV studies are blind to the most deeply embedded phases. The stars must already be clearing their surroundings to be observable at these wavelengths, meaning that these methods trace when clusters become visible, not when star formation starts or gas dispersal begins. The resulting timescales are therefore biased toward later, more exposed stages.

Infrared (IR) observations fill part of this gap. Dust emission in the mid- and far-IR can trace embedded clusters before their stellar populations emerge in the optical or UV. When individual stars are resolved, their ages provide constraints on how long dust and gas remain detectable \citep[e.g.,][]{getman_age_2014}. Using this approach, studies with \textit{Spitzer} and \textit{Herschel} generally recover longer embedded lifetimes of $\sim3-5$ \Myr \citep[e.g.,][]{morales_stellar_2013, battersby_lifetimes_2017, kim_duration_2021, chevance_pre-supernova_2022}, reflecting their sensitivity to earlier stages of cluster formation.

However, IR observations face their own challenges. Dust emission is not exclusive to embedded clusters; contamination from field stars and unrelated star-forming regions can lead to false positives \citep{ascenso_embedded_2018}. Without high-resolution imaging and multi-wavelength context, it’s difficult to determine whether a dusty region truly corresponds to an embedded system \citep{johnson_panchromatic_2022}.

The capabilities of JWST are beginning to resolve this ambiguity. For the first time, we can spatially disentangle embedded structures in external galaxies, simultaneously tracing both stellar content and dust emission at sub-parsec resolution \citep[e.g.,][]{lee_phangs-jwst_2023, rodriguez_phangs-jwst_2023, rodriguez_tracing_2025}. In particular, polycyclic aromatic hydrocarbons (PAHs) serve as a powerful tracer of cluster emergence, illuminating photodissociation regions (PDRs) as UV radiation escapes young stars \citep{rodriguez_phangs-jwst_2023, whitmore_phangs-jwst_2023, knutas_feast_2025}. These signatures enable detailed mapping of the de-embedding process; but critically, they still require stellar ages to anchor timescales.

While there is general consensus that the de-embedding process takes less than 5 \Myr \citep{portegies_zwart_young_2010,schinnerer_molecular_2024}, the process is complex. Importantly, the interpretation of this timescale varies depending on the observable—whether it is CO, dust, PAH, or H$\alpha$ emission. Different tracers probe different phases of feedback and dispersal \citep{kruijssen_uncertainty_2018}, and different techniques emphasize different parts of the sequence \citep{leisawitz_co_1989}. Moreover, mounting evidence suggests that the embedded lifetime may not be fixed, but instead varies with the mass and environment of the parent cloud \citep{mcquaid_timescales_2024, leroy_cloud-scale_2025, knutas_feast_2025}.

Theoretically, it is likely that the lifetime of the embedded phase is closely tied to the formation and evolution of massive stars.
Winds and radiation from massive stars dominate the stellar feedback, so the timing of their formation is likely to be coupled to the timing of gas removal \citep[e.g.,][]{kim_duration_2021, lewis_early-forming_2023}.
However, measuring these formation timescales directly is extremely challenging, with the largest source of uncertainty being how stellar ages are measured \citep{getman_age_2014}. Additionally, it is unclear when embedded massive stars can be detected individually as protostars, or how long it takes for high mass stars to form \citep[e.g.,][]{backs_properties_2024}.  

To address these challenges, simulations provide a valuable alternative. With complete knowledge of stellar properties and gas dynamics, simulations can isolate the physical mechanisms that regulate embedded lifetimes and test how different tracers respond to them. As a proven laboratory for testing star formation models, the recent increase in high performance computing has allowed simulation studies to make large leaps in the models of cluster growth \citep[e.g.,][]{karam_modelling_2022, lahen_griffin_2020}, embedded cluster morphology, and N-body dynamics \citep[e.g.,][]{cournoyer-cloutier_early_2023}. 

One of the simulation suites pushing the envelope in resolution is STARFORGE \citep{grudic_starforge_2021}. With a mass resolution of 10$^{-3}$M$_{\odot}$, STARFORGE fully models individual star formation, incorporating stellar processes including all major forms of feedback along with global properties of gravity and magnetic fields to create a self-consistent model of star formation. 

Recent analysis of the STARFORGE simulations has measured the gas removal timescales for several different initial conditions, finding that the gas removal occurs rapidly, taking between 0.7 and 1.7 \Myr and around 0.5 cloud free-fall times \citep{farias_stellar_2024}. In this work, we build on these measurements, focusing primarily on how this gas removal impacts the embedded cluster's observability in a variety of ways. 

Specifically, we leverage the high resolution of the STARFORGE simulations to investigate the timescales that proto-clusters remain embedded. To do this, we investigate the governing processes of gas removal, the timing and duration of these mechanisms, the drivers of these processes, and the observable signatures of different stages of this process. We perform this measurement for the stellar region as a whole, considering the global ensemble if properties, as well as for individual stars.

This paper is structured as follows. We first introduce the STARFORGE simulations in Section~\ref{sec:sf_sims}. We then describe our methodology for measuring stellar properties (Section~\ref{sec:age_determ}), and extinction (Section~\ref{sec:measure_av}), and present the evolution of embeddedness in Sections~\ref{sec:clst_bed}, \ref{sec:star_bed}. We then present different  wavelength observations of the simulation in Section~\ref{sec:observations}, highlighting timescales in Section~\ref{sec:dom_lum_timescales}. We then discuss the implications of these measurements in Section~\ref{sec:contextualizing}. 

\section{STARFORGE Simulations}
\label{sec:sf_sims}

We analyze a set of 3D radiative magnetohydrodynamical (MHD) STARFORGE simulations \citep{grudic_starforge_2021}, which account for all key ingredients needed to describe the dynamics of gravity, MHD, chemistry, radiative cooling, stellar dynamics, and all forms of stellar feedback. Briefly, STARFORGE uses the \texttt{GIZMO} code \citet{hopkins_new_2015}, and the ``full'' STARFORGE physics setup of \citet{grudic_dynamics_2022} which includes numerical methods for sink particles, stellar feedback, and evolution, as described in \citet{grudic_starforge_2021}. These simulations have sufficient mass resolution to resolve the stellar initial mass function (IMF) down to a $\sim 0.1 M_\odot$, enabling resolution of individual protostellar core-collapse, accretion, and feedback flows.

STARFORGE solves the ideal MHD equations using the \texttt{GIZMO} Meshless Finite Mass MHD solver \citep{hopkins_accurate_2016}, with a constrained-gradient $\div \mathbf{B}$-minimization scheme \citep{hopkins_constrained-gradient_2016}, and a fixed Lagrangian mass resolution typically set to 10$^{-3}$M$_{\odot}$. The time-dependent, frequency-integrated radiation hydrodynamics equations are solved by the \texttt{GIZMO} mesh-free finite-volume solver assuming the M1 closure \citep{hopkins_numerical_2019, hopkins_radiative_2020}, with a reduced speed of light of 90 km s$^{-1}$, in five frequency bands ranging from the Lyman continuum to the far IR. The different bands couple to matter through photoionization, photodissociation, photoelectric heating, and dust absorption, accounting for photon momentum. Dust, gas, and radiation temperatures are evolved independently, and self-consistently with the IR band of the RT solver, while radiative heating and cooling from all major molecular, atomic, nebular, and continuum processes using the cooling module of the FIRE-3 simulations \citep{hopkins_fire-3_2023}. 

In STARFORGE, a gas cell forms a sink particle once it satisfies a list of criteria designed to identify the centers of runaway collapse (see \citealt{grudic_starforge_2021}). The newly formed sink particle can then accrete gas based on the criteria of \citet{bate_modelling_1995}, which include the gas gravitational boundedness, its proximity to the sink particle, angular momentum, and density. Once a particle has become a sink, its equation of motion includes only gravity and the back-reaction of local momentum conservation during accretion. Binary and higher-order multiple systems can form through the dynamics of distinct sinks \citep[e.g.,][]{guszejnov_effects_2023}, but we do not resolve multiplicity for separations below $\sim20$ au and each sink is treated as a single star for the purposes of this work.

Each sink hosts a single model protostar and an internal gas ``reservoir''. Resolved accretion moves gas from the simulation domain into the sink by merging the discrete gas cells, and unresolved accretion then transfers mass from the reservoir onto the protostar, providing a continuous accretion rate in this regime \citep{,grudic_starforge_2021}\footnote{See Figure 4 in \citet{grudic_starforge_2021} for a complete picture of the sink, star, reservoir, and resulting feedback.}. The protostar, and its properties such as luminosity and radius evolves through standard pre–collapse and protostellar phases to the main sequence \citep{offner_effects_2009}. We approximate the emergent spectrum as a blackbody with
$T_{\rm eff} = 5780\,{\rm K}\,(L_*/L_\odot)^{1/4}(R_*/R_\odot)^{-1/2}$. Protostars can continue to accrete after reaching the main sequence, which we discuss further in Section~\ref{sec:age_determ}. 

Feedback is tied directly to this evolving protostar: protostellar outflows scale with the accretion rate, while radiative feedback and line–driven winds scale with the stellar mass and luminosity \citep{grudic_starforge_2021,grudic_dynamics_2022}. There is thus no single discrete ``onset'' of feedback; mechanical feedback is present from the moment a sink forms and radiative/wind feedback strengthen as massive stars grow. Self–gravity is solved on a finite, non–periodic domain whose size is $\sim$10 times the initial cloud radius, so the cloud evolves as an effectively isolated object rather than in a periodically replicated volume.

In this work, we primarily focus on a fiducial simulation\footnote{named: M2e4\_R10}, which begins with a molecular cloud of uniform density. The cloud has a mass of 2000 M$_{\odot}$, a radius of 10\pc, in a 100\pc periodic box filled with an ambient medium 1000$\times$ less dense than the original cloud. The flow is initialized as a Gaussian random velocity field with a $k^{-2}$ power spectrum and scaled to obtain an initial virial parameter of 2, such that the cloud is marginally bound. The initial magnetic field is uniform and scaled so that the magnetic pressure is 10\% of the turbulent ram pressure \citep{miville-deschenes_physical_2017}. The mass resolution is 10$^{-3}$ M$_{\odot}$ for gas cells, and 10$^{-4}$ M$_{\odot}$ for protostellar jets and stellar wind gas cells. In section \ref{sec:sob_ts}, we also analyze a second simulation\footnote{named: M2e5\_R30}, which has a mass of 20000 M$_{\odot}$, and a radius of 30\pc but the same governing physics as the fiducial model. 

\section{Quantifying the Properties of Embedded Stars in STARFORGE}
\label{sec:methods}

In our subsequent analysis, we analyze the time evolution of the properties of the proto-cluster stars, as they form and move out of the embedded phase. However, doing so requires quantifying a number of stellar properties in a consistent fashion. For stellar masses, it is simple; we chose to associate each star with its eventual final mass after accretion ends, even when considering it at an earlier time in its evolution. However, for age and extinction, there is more nuance in the definitions, and we discuss the specific choices in the subsequent subsections.

\subsection{Quantifying Stellar Ages}\label{sec:age_determ}

As the simulated cloud evolves, the initial turbulent motions seed dense clumps and filaments. These high density clumps and filaments have shorter dynamical times, which can accelerate star formation locally. As a result, there can be a significant age spread for the formation times of protostars in the cluster. Moreover, protostars can continue accreting gas from the cloud, making the exact ``formation time'' of the final star ambiguous.

\begin{figure}[t]
    \centering
    \includegraphics[width=0.47\textwidth]{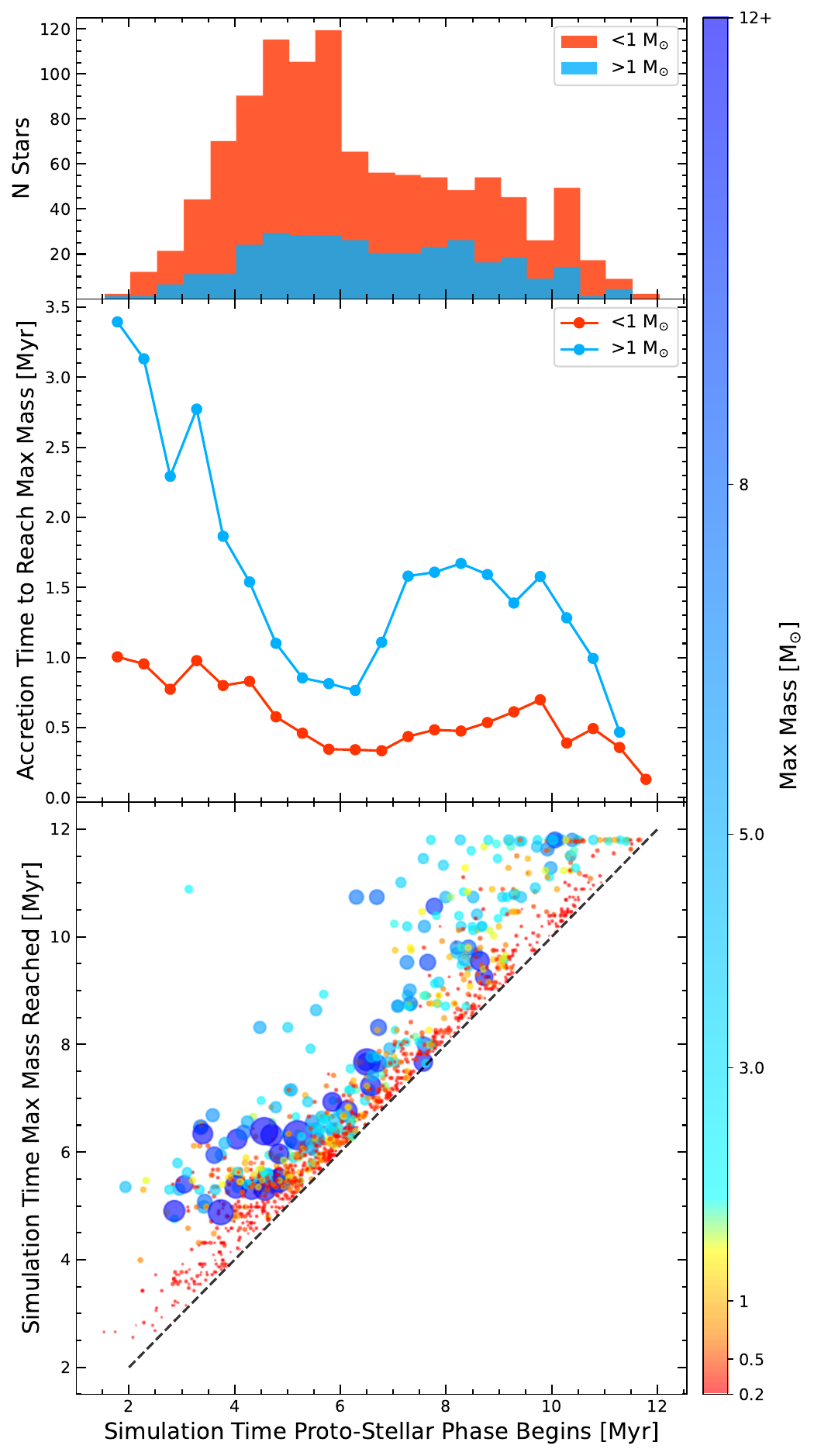}
    \caption{The simulation time where a sink particle begins to accrete and how long it takes for that particle to reach its maximum mass, which we take to be the zero-point of a star's age. The top panels shows the distribution of stellar formation times, split into two bins, one for stars below 1 M$_\odot$, and one for stars larger than 1 M$_\odot$. The middle panel then shows the average time it takes for the stars in these two bins to achieve their maximum mass. In both bins, but specifically for the larger stars, the accretion time depends on simulation time, which encodes the gas density. The bottom panel then shows each individual star, colored and sized by its final mass, where the y-axis shows the simulation time where each star reaches its maximum mass. The accretion time is then vertical distance above the dashed one-to-one line.} 
    \label{fig:age_determ}
\end{figure}

We demonstrate these effects and show the resulting age spread in Figure~\ref{fig:age_determ}. In the bottom panel, for each star we compare the simulation time when each protostar begins it's protostellar phase (on the X-axis) to the time when the star reaches its final mass (on the Y-axis), color-coded by the final mass. The histogram in the upper panel shows the age distribution for the start of the protostellar phase, for all stars (red) and for the subset of more massive stars ($>1\msun$; blue) \footnote{We explored further binning the stars, but found that all bins above 1 M$_{\odot}$ behaved the same.}. The middle panel shows the total duration of the protostars' mass-accretion stage for the same mass bins. 

The density distribution histograms of stellar age, and range of points in the upper panel of Figure~\ref{fig:age_determ} clearly demonstrates the extended nature of star formation in the cloud. Protostars begin to form within 2 Myrs of the start of the simulation, and while the formation peaks at $4\mbox{--}6$ \Myr, new protostars continue to form until well after 10 Myr, such that over $\sim$40\% of stars form after 6.5 \Myr, and the rms in the stellar formation times is 6.7 \Myr.

In addition to the scatter in proto-star formation times, the accretion time to reach the final mass can be significant.
In the bottom panel of Figure~\ref{fig:age_determ}, for each point, this accretion time is the integrated distance between the star, and the black dashed line, while in the middle panel, this time is shown on the y-axis. Of particular interest is the accretion time for stars move massive than 1\msun, which may take upwards of 3 Myr to reach their final masses \citep{grudic_dynamics_2022}, depending on the availability of the gas supply. At early times, gas is ample, leading to longer accretion times for early-forming stars, on average. 

We note, however, the accretion time does not decrease monotonically with time. This variation is driven by substructure in the cloud, such that the gas supply varies as a function of position. Some parts of the cloud may collapse at later times, in regions with less local feedback and thus larger local gas reservoirs. These local gas pockets can fuel on-going accretion, explaining the second period where accretion rates for stars more massive than 1 M$_\odot$ rise above 1 Myr between 7-10\Myr. 

\subsubsection{Stellar Age Definition}
Because the stars in our simulations form over an extended period, we cannot assume a single, shared formation time. Instead, we define each star’s ``zero age'' as the moment it reaches its maximum mass. This choice avoids two problematic alternatives: the time when a sink particle first forms, which is highly stochastic due to varying accretion times; or the time the star reaches the main sequence, as massive stars can continue to accrete after reaching the main sequence. Specifically, massive stars in our simulations continue to accrete for a median of 0.27 Myr after reaching the main sequence, and in some cases for as long as 2 Myr, meaning that main sequence arrival does not mark the end of mass growth or feedback onset. While our definition lacks a direct observational analog, it more closely reflects the physical endpoint of a stars formation and provides a consistent, replicable anchor which uniquely describes the timescales probed in this study.


\subsubsection{Exploring Stellar Age and Location}
\label{sec:location}

As the system dynamically evolves throughout the simulation, the location of both the stars, and the gas are constantly changing. As discussed in Section~\ref{sec:age_determ}, we see a large age spread as different sub-clumps within the cloud collapse and form stars. In this section, we quantify the location of the stars with respect to their age.

\begin{figure*}[ht]
    \begin{interactive}{js}{Figures/interactive_gas_and_plot.html}
    \centering
    \includegraphics[width=0.97\textwidth]{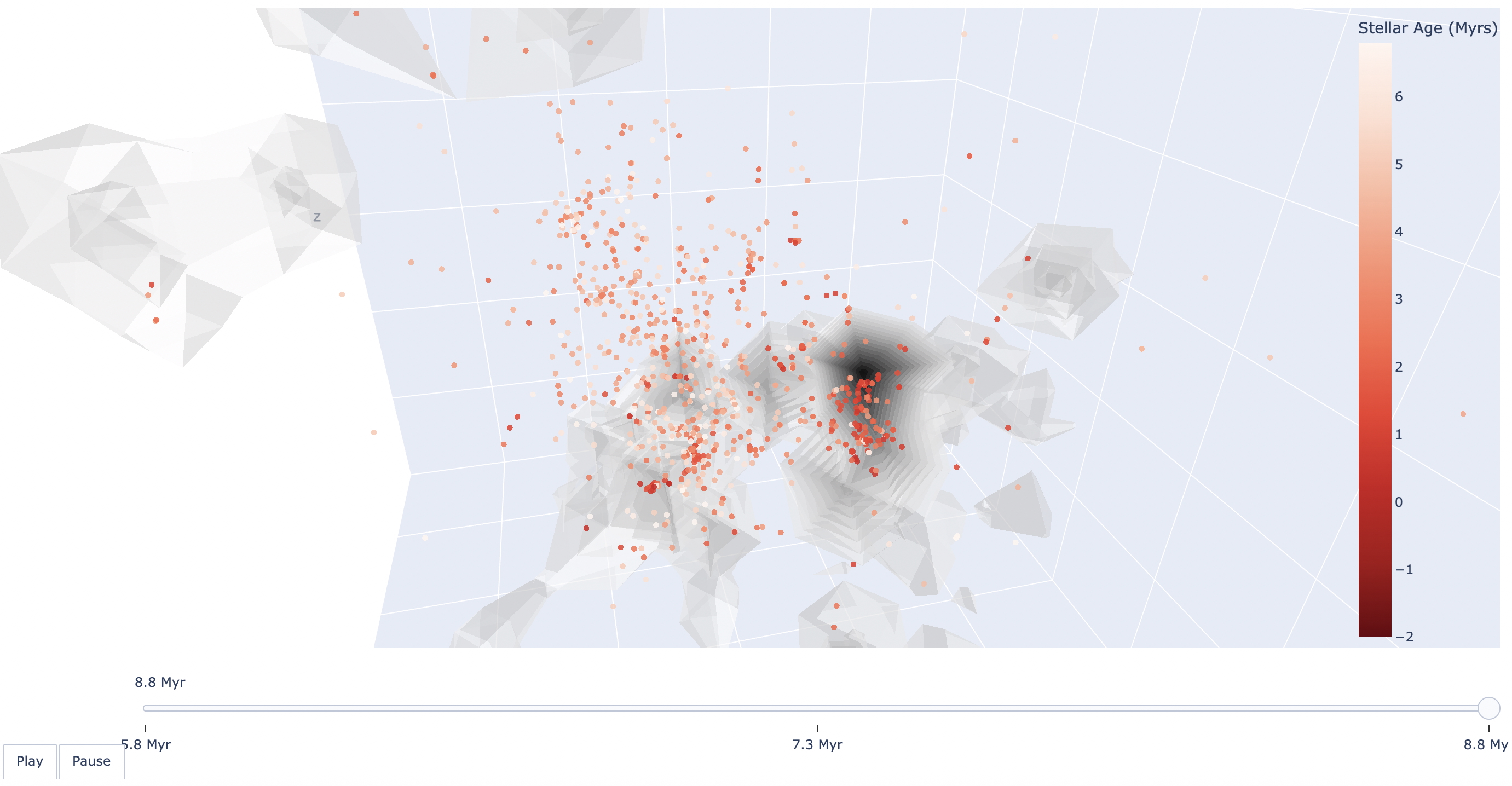}
    \end{interactive}
    \caption{Interactive figure showing stellar and gas locations at three simulation snapshots, mapped to the middle panels of Figure~\ref{fig:sim_av_vs_age}. The red scatter points indicate stellar locations, where the points are colored by stellar age as defined in Section~\ref{sec:age_determ} where negative ages (darker red colors) are stars that are still undergoing accretion. The gray contours depict the gas density at that location. The white lines on the grid mark 5 pc. The still snapshot shown is at 8.8\Myr, and demonstrates that the ongoing star formation is for a localized sub-clump in the right part of the image. The interactive version of the plot includes 5.8\Myr, as well as 7.3\Myr, where readers can rotate around simulation space and zoom in/out. }
    \label{fig:interactive}
    
\end{figure*}

Throughout the simulation, the stars form in meaningfully different locations. Initially, the cloud spans 20\pc and then collapses, which leads to stars forming centrally. However at later times, the cloud expands and formation occurs on the outskirts. Therefore, the median formation location for stars that form before 6.5 \Myr is 4.1 pc away from the median location of formation for stars after 6.5 \Myr. This distance is also demonstrated in Figure~\ref{fig:interactive} where the youngest stars are noticeably offset from the older stars. 

The clumpy nature of the stars and gas is demonstrated in the interactive Figure~\ref{fig:interactive}. Shown in red are the 3 dimensional locations of stars, colored by their age (as defined in Section~\ref{sec:age_determ}), where older stars are whiter, and the youngest stars still undergoing accretion are darker. We can clearly see that the youngest stars are tightly clustered in the bottom right corner of the still image. Readers are encouraged to rotate around simulation space and view the location of stars from several angles. 

In addition to the stellar locations, the gray contours show gas density. At 8.8 \Myr, we see small pockets of gas on the outskirts of the stellar association. Importantly, the location of the youngest stars directly corresponds to where the gas is the densest. This correlation between pockets of gas and newly formed stars demonstrates the sub-clump nature of the star formation in the model, and explains the large spread in stellar ages discussed in Section~\ref{sec:age_determ}, where parts of the cloud collapse at different times. 

One way to describe this correlation quantitatively is to measure the median distance to all other stars. This stellar density also can be compared between frames of the interactive Figure~\ref{fig:interactive}. Stars that form before 6.5 \Myr are tightly clustered in physical space, with a median distance to other stars of 3.5\pc~. However, when we consider stars that form between 6.5 \Myr, and 10 \Myr, the median distance to other stars almost doubles to 6.8\pc~. This latter increased distance to other stars is consistent with a more-distributed mode of star formation that ensues when the already expanding cloud also begins disruption. This increased distance is also clearly seen in the interactive Figure~\ref{fig:interactive} by adjusting the slider between times; the stars become less tightly clustered as time progresses.

\subsection{Quantifying Extinction} \label{sec:measure_av}

Quantifying the extinction associated with individual stars, or with the proto-cluster as a whole, is a critical part of assessing ``embeddedness''. However, like age or mass, there are a range of possible choices one could conceivably make. 

We start by assuming that gas and dust are well-mixed and that intrinsic dust properties are uniform, so that dust extinction is determined solely by the gas column density along the line of sight. We compute the gas column density by summing the column densities of gas elements along each stellar line-of-sight. We then convert to the optical extinction $A_{\rm V}$ using the conversion $A_{\rm V} = 4.52 \times 10^{-22} \mathrm{cm}^2\,N_{\rm H}$ from \citet{guver_relation_2009}, which was derived via x-ray measured column densities with respect to optical extinctions.
We take $N_{\rm H}$ to be the number column density of H nuclei (i.e. not distinguishing between ISM phases).

The extinction, $A_{\rm V}$, to an individual star depends strongly on the complex, 3-dimensional structure of the surrounding gas cloud. As such, the extinction is a strong function of viewing angle, and any one line-of-sight measurement of $A_{\rm V}$ is not necessarily representative of the overall embeddedness. 

We therefore characterize the ``typical’’ extinction to a star by recalculating the line-of-sight extinction, $A_{\rm V_{\theta,\phi}}$, for 60 unique viewing angles. These directions are distributed quasi-uniformly over the sphere using a Fibonacci lattice, with step sizes in spherical coordinates where $\phi = \pi (3 - \sqrt{5})$ and $\theta = \phi \cdot i$, where $i$ indexes the viewing direction. This yields a well-sampled distribution of extinctions for each star at a given timestep. We adopt the median of these values, $\widetilde{A_{\rm V}}$, as a representative metric for describing a star’s overall level of embeddedness. This analysis is repeated every 8th simulation snapshot (i.e., approximately every 0.2 Myr), which we find to be fine-enough sampling to accurately describe the simulation for the present analysis.

\begin{figure*}[ht!]
    \centering
    \includegraphics[width=0.95\textwidth]{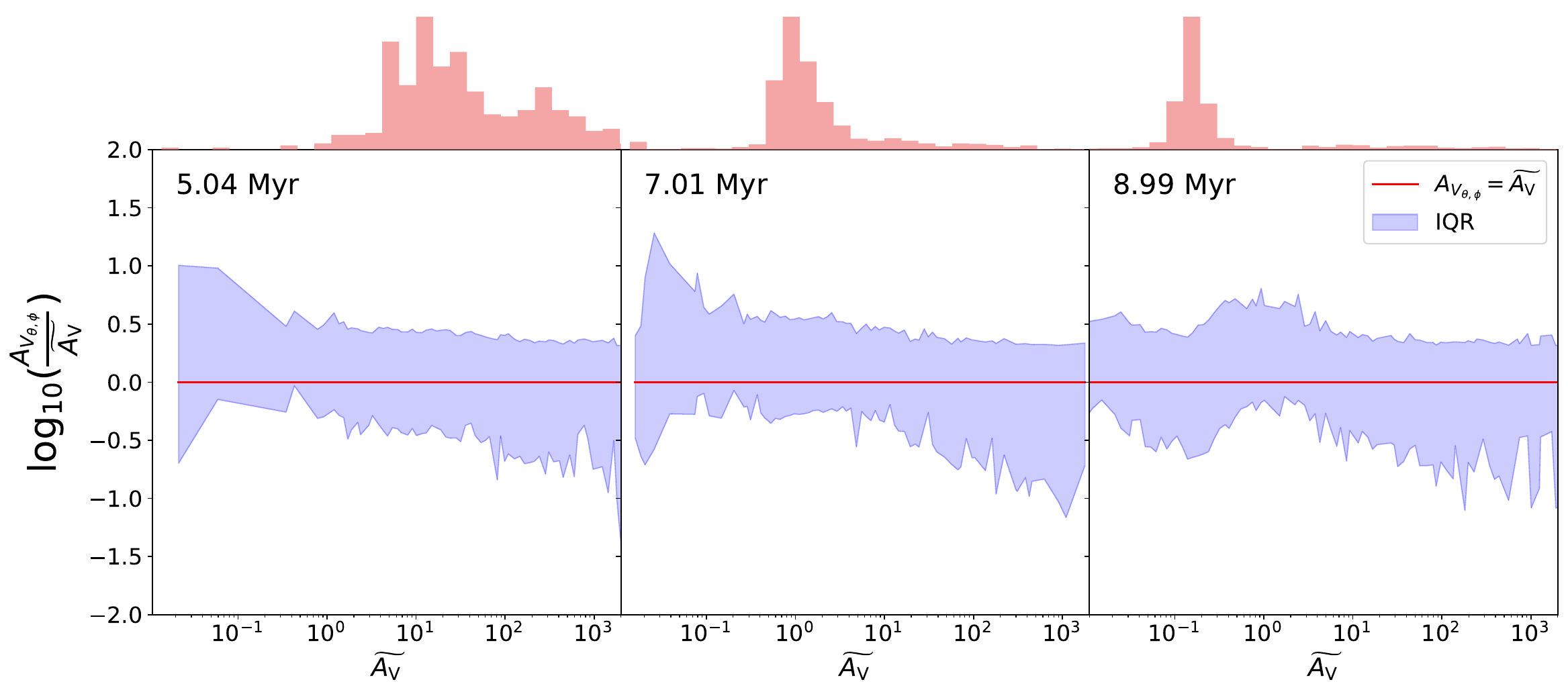}
    \caption{Interquartile ranges (IQR) for the log ratio of stellar $A_{\rm V}$s at different viewing angles, A$_{\rm V_{\theta,\phi}}$, and a function of the median extinction ${\widetilde{A_{\rm V}}}$, shown for three different times in the simulation. Shown in red is the one-to-one line where A$_{\rm V_{\theta,\phi}}$ is equal to ${\widetilde{A_{\rm V}}}$, while the blue shaded region shows inner quartile range. The histograms above are the number density of stars at a given ${\widetilde{A_{\rm V}}}$ at that time, showing how many stars are contributing to the IQR. }
    \label{fig:av_spread}
\end{figure*}

Figure~\ref{fig:av_spread} shows the resulting distribution of extinctions at different viewing angles, for three different time steps in the simulation. The $x$-axis shows the value of the median extinction ${\widetilde{A_{\rm V}}}$, in a running bin space evenly in log space. The $y$-axis shows the interquartile range (in blue) of the values of the individual orientation measurements $A_{\rm V_{\theta,\phi}}$, normalized by the median. The total number of stars at each median extinction ${\widetilde{A_{\rm V}}}$ is plotted as a red histogram above the plotting window, and shifts from high ${\widetilde{A_{\rm V}}}$ to low as the gas in the cluster is consumed or expelled.  

In the left most panels of Figure~\ref{fig:av_spread}, the interquartile range is relatively constant in log space, which is consistent with the log-normal nature of most ISM density fields. We find that it typically spans around 0$\pm$0.5 in the log, independent of the median extinction, or the timestep.  

We note, however, the distribution shifts from being heavy-tailed towards high densities for weakly embedded stars (low ${\widetilde{A_{\rm V}}}$), to being skewed towards low densities for highly embedded stars (high ${\widetilde{A_{\rm V}}}$). This tail is largely an artifact of ``reversion to the mean''. Statistically, when a star has an extreme level of ${\widetilde{A_{\rm V}}}$, it is far more likely that there will be a tail to more ``typical'' levels of extinction for the association, which explains the characteristic asymmetry.

The behavior of the distribution of $A_{\rm V_{\theta,\phi}}$ is somewhat more complex at late times (rightmost panel of Figure~\ref{fig:av_spread}). Rather than the skew shifting steadily from low to high ${\widetilde{A_{\rm V}}}$, instead, the characteristic change in skew occurs twice, once at low median extinctions and again at high ${\widetilde{A_{\rm V}}}$. This behavior likely reflects the presence of a residual dense substructure within a cloud that has largely been dissipated by feedback. The bulk of the stars have median extinctions of ${\widetilde{A_{\rm V}}}\sim0.2$, but there is a long tail with extinctions greater than 1. Specifically, there are 1004 stars with an ${\widetilde{A_{\rm V}} < 1}$, and only 170 with an ${\widetilde{A_{\rm V}} > 1}$. The overall distribution of $A_{\rm V_{\theta,\phi}}$ is therefore the sum of two distributions -- one that is characteristic of stars that live in the surviving dense substructure, and one that is characteristic of the stars that occupy the largely evacuated cloud. For the stars in the tail of this distribution, the IQR is therefore more stochastic since it is based on a small number of sources. 

\section{The Evolution of Embeddedness During Formation} \label{sec:results}

With our extinction measurements in hand, we can now track how stars transition from being deeply embedded to fully exposed over time. In the following section, we apply this framework to quantify the changing embeddedness of both the stellar population as a whole and individual stars. We begin by characterizing the global evolution of embeddedness over time, then identify the key physical mechanisms that drive this evolution. We also explore embeddedness for individual stars, and investigate the strong mass dependencies in how long stars remain embedded. Finally, we examine how the overall timescales of embeddedness vary across two clouds of different mass.

\subsection{The Overall Evolution of Embeddedness}\label{sec:clst_bed}

The expectation from models of stellar feedback is that stars form from dense gas within dusty molecular clouds, but that the evolution of these stars eventually destroys and/or removes the molecular gas, through the cumulative impacts of jets, winds, and radiation, and eventually supernova. Within this framework, stars that form early should be deeply embedded in the dusty gas, leading to high ${\widetilde{A_{\rm V}}}$. In contrast, by the end of the cloud's evolution, most stars occupy regions that are free of gas, and thus should be visible along largely clear lines-of-sight. The only exception would be any newly formed stars, which may still occupy dense dusty sub-clumps that happened to survive to late times.

\begin{figure*}[htp]
    \centering
    \includegraphics[width=0.97\textwidth]{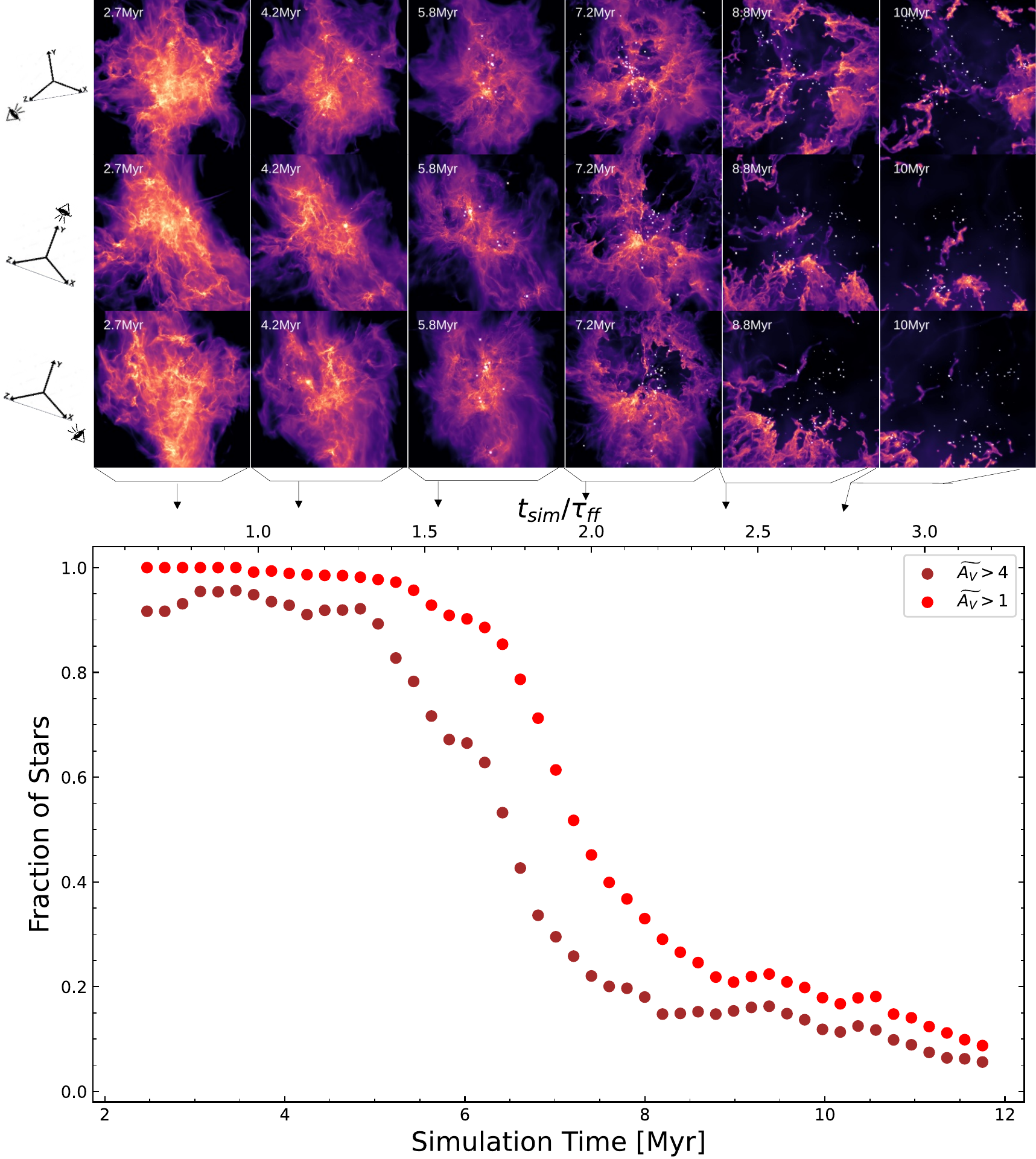}
    \caption{{\it Top}: Snapshots from six simulation times each box covering 30\pc: different rows correspond to three different viewing angles. Gas surface density is shown in the color map, and stars rendered with a standard HST PSF. Stellar magnitudes are determined by considering the line of sight gas column density. {\it Bottom:} Scatter plot showing the fraction of stars with a ${\widetilde{A_{\rm V}}}$ greater than 4 (brown), and 1 (red) as a function of simulation time, where the median is defined by 60 different viewing angles. The stars in the formation event emerge from their natal cloud relatively quickly, with a steep decline in the fraction of stars that are highly extincted occurring over a $\sim$1 Myr period at 6 \Myr. It is also important to note that this model does not produce a gravitationally bound stellar cluster. Rather, the association disperses entirely by 10 Myr, which can be seen in the right most panel of simulation snapshots. }
    \label{fig:sim_av_vs_age}
\end{figure*}

We can see this evolution in the top panels of Figure~\ref{fig:sim_av_vs_age}, which plots snapshots of the gas and stars in the fiducial STARFORGE simulation as a function of increasing time (from left to right, spanning from 2.7-10\Myr) and three different viewing angles (from top to bottom); the gas density is shown in the color map along with individual stars (in white) that are convolved with the HST point spread function. In this sequence of images, one can see the expected transition from gas-rich with few visible stars ($<$5\Myr), through a transition period with a mix of stars occupying evacuated regions within the cloud and dense gas that is likely to still be forming stars ($\sim$5-8\Myr), to finally a late stage where most of the molecular cloud has been destroyed, leaving the majority of stars visible from most viewing angles, outside of a few residual dense gas clumps ($>$8\Myr).

We quantify this evolution in the bottom panel of Figure~\ref{fig:sim_av_vs_age}, where we show the fraction of stars that are deeply embedded (${\widetilde{A_{\rm V}}} >4$~mag; brown) or weakly embedded (${\widetilde{A_{\rm V}}}>1$~mag; red), as a function of time. These limits were chosen because in solar-metallicity, photo-dissociation models, $A_V\sim1$~mag is the approximate threshold for forming molecular hydrogen \citep{franco_molecular_1986, krumholz_atomic--molecular_2009, sternberg_h_2014}, where $A_V\sim4$~mag is a threshold within which carbon recombination takes place, and corresponds to HCN line emission \citep[e.g.,][]{hollenbach_dense_1997, kauffmann_molecular_2017}. The small vertical arrows connecting the top panels to the bottom plot indicate the times corresponding to the snapshot images. The times are given both in terms of the time since the start of the simulation ($t_{sim}$; bottom axis), and $t_{sim}$ scaled by the free-fall time of the initial gas cloud ($\tau_{ff}$; top axis, calculated as $\tau_{ff}=3.7$\Myr). Every point on the bottom plot includes all stars formed at or before $t_{sim}$, not just the stars forming at that particular timestep.

Figure~\ref{fig:sim_av_vs_age} shows that stars emerge from their natal cloud relatively quickly. Specifically, there is a steep decline in the fraction of stars with a ${\widetilde{A_{\rm V}}}>1$ occurring over a rapid $\sim$2.5 \Myr period. At a simulation time 6~\Myr, 90\% of stars have a ${\widetilde{A_{\rm V}}}>1$, but only 2.5~\Myr later, fewer than 25\% of stars have even this relatively modest level of extinction. 

Although the overall emergence of stars is rapid, a reservoir of dense gas persists for much longer, even after half the stars have ${\widetilde{A_{\rm V}}}>1$ (at $t_{sim}=7.3\,\Myr$; see the top panel of Figure~\ref{fig:sim_av_vs_age}).This reservoir consists of dense pockets of gas that still continue to form stars. While some viewing angles (such as the bottom row at late times) have sufficiently cleared enough local gas to open an unobstructed viewing window, other viewing angles (such as the top row) still have some lines of sight where there is some degree of gas for nearly all the stars. 

From the last two columns of Figure~\ref{fig:sim_av_vs_age}, where nearly all stars are largely un-embedded, it is also apparent that the simulation does not produce a gravitationally-bound stellar cluster. Instead, when the gas disperses, so do the stars, due to the low star formation efficiency of the system \citep[e.g.][]{tutukov_early_1978,hills_effect_1980}, leaving no long-lasting remnant of the star formation. Even as soon as $11\,\Myr$ after the initial collapse, while small pockets of gas are still producing a handful of stars, the stars from this event are not meaningfully coupled in any kind of longer-lived quasi-bound structure. 

Also shown on the secondary axis of the scatter plot is the time in units of the cloud free-fall time $\tau_{ff}$. We can see that the ${\widetilde{A_{\rm V}}}$ first begins to decrease around $\sim1.5$ $\tau_{ff}$. We further discuss the timescales driving the emergence in the following subsections. 

\subsection{Drivers of Embeddedness Evolution}
\label{sec:Drivers}

Empirically, the rapid emergence of stars from a dusty gas cloud is quite clear from the plots in Figure~\ref{fig:sim_av_vs_age}. Broadly, this change can be ascribed to ``stellar feedback'', but exactly what aspect of feedback drives the evolution? Is it early jets, which deposit kinetic energy into the gas? Is it the momentum deposition from stellar winds? Is it radiation pressure and/or dissociation from ultraviolet photons? All of these processes are potentially important in young star-forming regions --- as supported by an extensive literature examining these questions theoretically and observationally \citep[e.g.,][]{krumholz_dynamics_2009,lopez_what_2011,lopez_role_2014,kim_modeling_2018,grudic_nature_2019,olivier_evolution_2021}, all of which are likely stronger for massive stars. 

A full analysis of feedback is well outside the scope of this paper \citep[e.g.,][]{neralwar_effects_2024}, but we can attempt to identify whether any of the above processes are associated temporally with the rapid changes in the median stellar extinction. We specifically look at quantities that trace the formation of stars in general (the total number of stars, and the total bolometric luminosity of all stars and protostars), and quantities that trace the subset of massive stars that are likely to drive cloud evolution (the total number of stars with masses $>8\msun$, and the number of ionizing photons). 

When calculating the above quantities, we: (1) count stars during their formation, rather than from the time they stop accreting mass, to better capture effects from protostellar feedback; (2) calculate the bolometric luminosity using the effective temperature of the star as defined by its mass and radius for its stellar evolutionary phase; (3) and calculate the number of ionizing photons by assuming a blackbody stellar SED and integrating over ionizing wavelengths shorter than 91.1 nm. We also plot the time evolution of these quantities normalized by their peak value over the simulation.

\subsubsection{The Accounting of High-Mass Stars}

There is an additional subtlety to consider when counting stars. Specifically, in most models of feedback, massive stars have a greater impact, due to their higher luminosities, harder spectra, and higher average accretion rates during the protostellar phase \citep[e.g.,][]{dale_modelling_2015,lewis_early-forming_2023}. We therefore choose to track low- and high-mass stars separately during their evolution. 

However, during their early evolution, future high-mass stars can have the same masses as a future low-mass star. In STARFORGE, the strength of the instantaneous protostellar feedback is proportional to the mass accretion rate, which depends on the current mass of the protostar, as well as the local gas supply. As such, the impact of protostellar feedback may not strictly depend on the final mass of the star, during the star's early evolution.  

\begin{figure*}[t]
    \centering
    \includegraphics[width=0.95\textwidth]{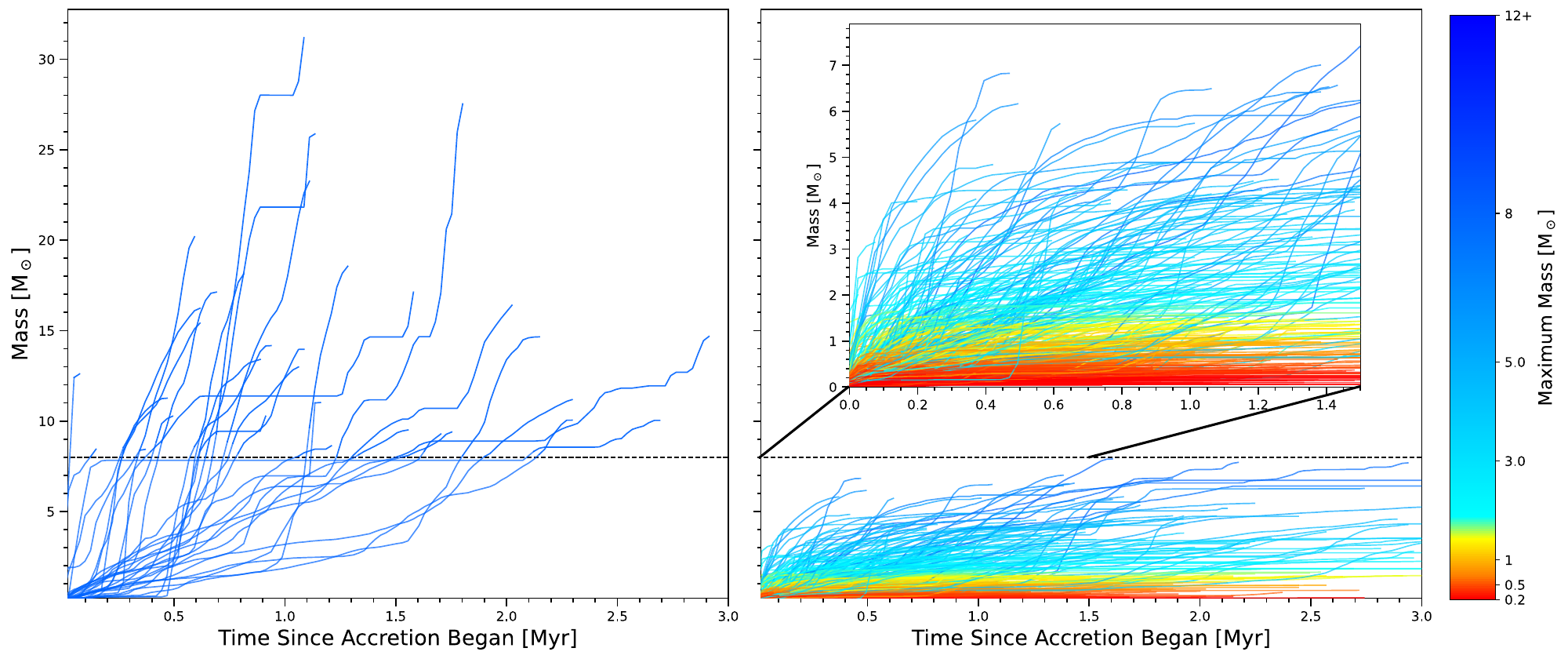}
    \caption{Accretion histories for stars in the simulation, colored by their final maximum mass, mapped to Figure~\ref{fig:age_determ}. Tracks for individual stars are only plotted until the point where the star reaches its maximum mass. The horizontal dashed line at 8 M$_\odot$ represents portions of the evolution considered in the massive star counter. Stars in the left panel are the 31 stars with a final masses greater than 8 M$_\odot$, while all other stars are in the right panel. We show an zoomed-in axis range in the right panel only going to 8 M$_\odot$ on the y axis, and 0 to 1.5 \Myr on the x-axis. }
    \label{fig:mass_accretion_vs_time}
\end{figure*}

We help clarify this subtlety in Figure~\ref{fig:mass_accretion_vs_time}, where we plot the mass evolution as a function of time for a sample of stars from our simulation, with stars with final masses greater than 8 M$_\odot$ in the left panel and all other stars on the right. Lines are color-coded by the eventual maximum mass of the star (according to Figure~\ref{fig:age_determ}), and the thick lines indicate the portion of the evolution where the mass is greater than 8 M$_\odot$, respectively. The accretion rate, proportional to the strength of protostellar feedback, is indicated by the slope of the tracks, such that steep, rapid increases in stellar mass will be associated with strong instantaneous feedback.  

For the majority of stars, the first $\sim0.3$ or so \Myr of each star's accretion in Figure~\ref{fig:mass_accretion_vs_time} looks similar, where massive stars may not immediately ``know'' that they are destined to be massive.
However, the eventually-massive protostars are then distinguished by having either: a rapid burst of accretion, or a systematically more sustained accretion rate over their lifetime -- as is needed to reach a higher final mass. It is this rapid and/or extended accretion that leads to higher protostellar feedback for massive stars.

We note however, that the high accretion rates seen for some massive stars in the left-hand panel (primarily at $<$5\Myr) are also observed for some of the low-mass stars in the zoomed-in version of the figure in the right panel. However, the rapid accretion that these stars experience is not as sustained as
for the massive stars in the left-hand panel. 
This systematic difference in the absolute accretion rate and/or its duration should lead to weaker protostellar feedback for the low-mass stars.  

In order to isolate the low-mass and high-mass phases of protostar evolution we count stars in two regimes: (1) any star that is currently less than 8 M$_\odot$, regardless of its final mass; and (2) any star that is currently greater than 8 M$_\odot$, regardless of its final mass.

\begin{figure}[ht]
    \centering
    \includegraphics[width=0.48\textwidth]{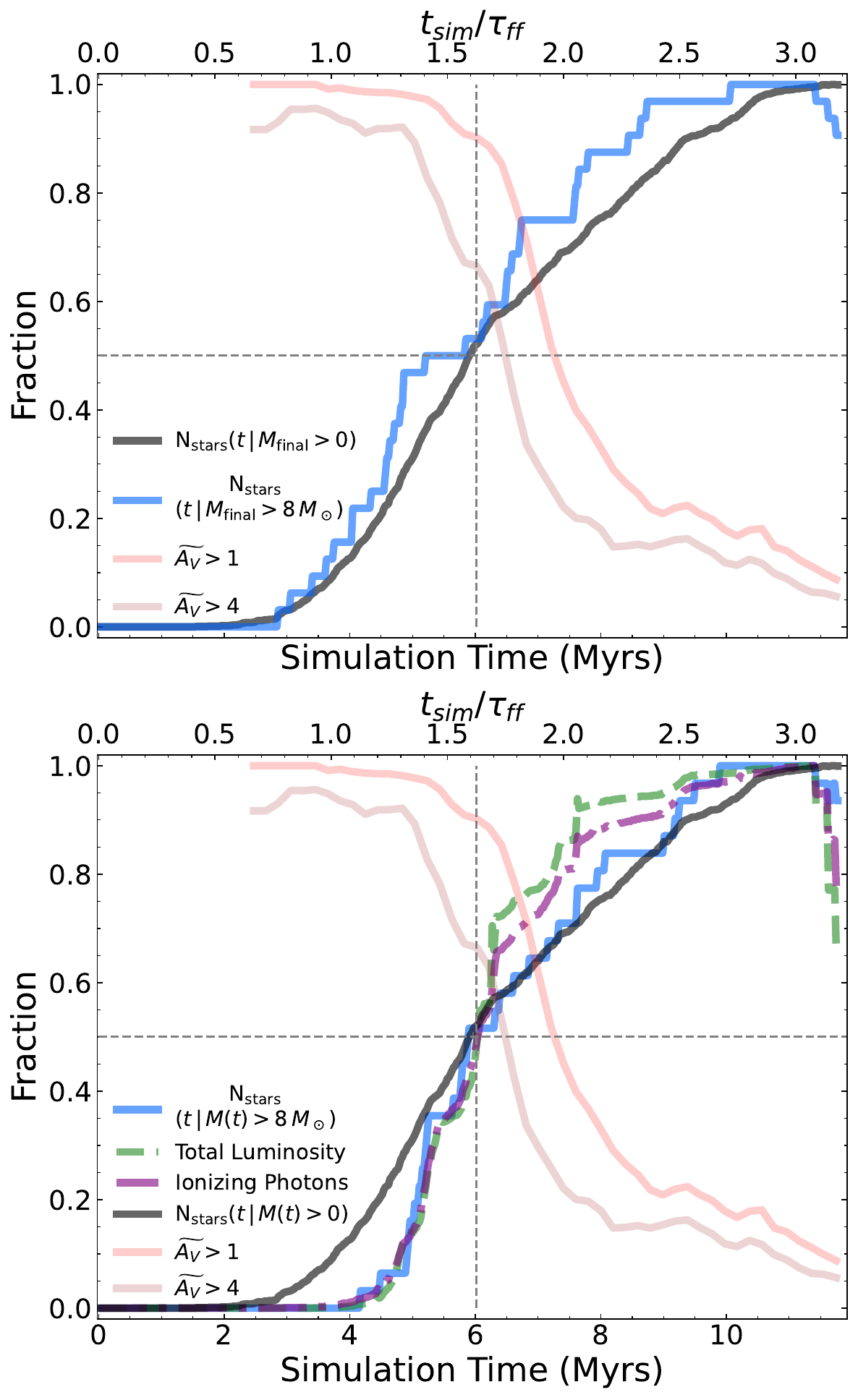}
    \caption{\textit{Top:} Distributions of the fraction of stars which have formed (black), as well as the fraction of stars larger than 8 M$_{\odot}$ (blue) which have begun accretion as a function if time. Both fractions are a ratio of the number at the current time step compared to the total amount that formed. Also shown is the fraction of stars with a ${\widetilde{A_{\rm V}}}$ greater than 1 (transparent red) and 4 (transparent brown) from Figure~\ref{fig:sim_av_vs_age} for comparison to the association embeddedness. In each panel, we show gray dashed lines at 50\% and at 6 \Myr, the point where ${\widetilde{A_{\rm V}}} >1$ begins to steeply drop. \textit{Bottom:} Bolometric stellar luminosity at each time step is shown in the dashed green line, and the number of ionizing photons is shown in dashed-dotted purple line, both normalized by the maximum value at any time. We see the fractional luminosity and the fraction of ionizing photons closely track the number of massive stars that have formed. Noticeably, each of the quantities converges to 50\% around the same time. }
    \label{fig:fid_counting}
\end{figure}
 
\subsubsection{The Time Evolution of Drivers}
With the definition of ``high mass" now settled, in Figure~\ref{fig:fid_counting} we compare the time evolution of the embeddedness to the time evolution of a variety of possible drivers of stellar feedback to look for possible correlations. In both panels, we plot the time evolution of the embeddedness, tracked using the fraction of stars that are heavily embedded (${\widetilde{A_{\rm V}}}>4$; brown) or lightly embedded (${\widetilde{A_{\rm V}}}>1$;red), reproduced from Figure~\ref{fig:sim_av_vs_age} and discussed in Section~\ref{sec:clst_bed}. 

In the top panel, we compare the change in the fraction of embedded stars to the fraction of stars which have begun their protostellar phase, for both stars with high final masses (M$_{final}>8\msun$; dark blue), and all stars regardless of their final mass (black line). We count massive stars from the time accretion begins until the time they explode and normalize the curves by the maximum number of stars in each category at any point during the simulation. Note that this normalization leads to a downturn in the fraction of massive stars at $\sim$11\Myr in Figure~\ref{fig:fid_counting}, when SNe begin. 

Comparing the number of stars to the embeddedness shows the expected impact of stellar feedback, namely that the stellar extinction (highly transparent red and brown lines) drops systematically as more stars form. 
However, a surprisingly large amount of star formation occurs before we see any clear decline in extinction. Specifically, the $\widetilde{A_V} > 4$ remains above 90\% until $\sim$4.7\Myr, and $\widetilde{A_V}>1$ remains above 90\% until $\sim$6.0\Myr, when both begin to decline more rapidly. Quantitatively, at $\sim$4.7\Myr, only 25\% of all stars have already begun forming. Perhaps more strikingly, close to 50\% of eventually massive stars have begun forming by $\sim$6.0\Myr, when 
there is a significant drop in the typical embeddedness.

\subsubsection{Only Considering Massive Stars Once they are Massive}

This unexpected result can be explained by considering how the stars are being counted. The top panel counts stars from the time they begin accretion, but it is clear from Figure~\ref{fig:mass_accretion_vs_time} that even ``eventually massive'' stars can exist upwards of 2 \Myr before they are meaningfully more massive than the other stars. As such, there is no appreciable difference between the blue and black curves in the top panel of Figure~\ref{fig:fid_counting}, and no guarantee that any of the stars contributing to the curves currently have high mass, early in the simulation.

In contrast, we see very different behavior in the bottom panel of Figure~\ref{fig:fid_counting}, where we choose to only consider massive stars once they surpass the 8 M$_\odot$ threshold (i.e., above the dashed line in the left panel of Figure~\ref{fig:mass_accretion_vs_time}). With this choice, we are only counting massive stars once they are massive enough to have had either, significant mass accretion rates (i.e., significant protostellar feedback), or prolonged accretion events, such that they have large luminosities.

With this different choice of how to define ``massive'', we see a different correlation with the $A_V>4$ embeddedness that tracks dense gas. In contrast to the top panel, the decrease in extinction now appears to be more tightly related to the onset of massive stars (at $\sim$4-5\Myr) rather than star formation in general (comparing the blue and grey curves). In other words, the drop in significant stellar embeddeding is driven by effects produced by stars as they reach a significant mass.

While the observed correlation strongly suggests that massive stars are driving the drop in extinction, there are multiple possible mechanisms at work. We consider one of these mechanisms in the bottom panel of Figure~\ref{fig:fid_counting}, where we assess the role of radiation, which is dominated by the population of hot massive stars. We plot the time evolution of the bolometric stellar luminosity (green) and the number of ionizing photons (purple), each normalized by their maximum value throughout the simulation\footnote{For reference, the maximum number of ionizing photons is $10^{48.09}$ photons, and the maximum stellar luminosity is $10^{5.92}$ L$_\odot$.}. 

As expected, both the luminosity and the number of ionizing photons increase along with the number of stars. Even more striking is the degree to which both quantities track the subset of massive ($>8$ M$_\odot$) stars, rising in lockstep until 6\Myr, which is roughly the time by which half of all stars (massive and otherwise) have formed, half the sight-lines are no longer highly embedded (transparent brown $\widetilde{A_V}>4$ curve), and the fraction of lightly embedded stars begins to decline (transparent red $\widetilde{A_V}>1$ curve). At later times, the total stellar luminosity and the number of ionizing photons grow even faster, presumably as a subset of stars reach masses higher than the 8 M$_\odot$ threshold used as the mass limit for the blue curve.

The impact of the most massive stars on the total luminosity is also seen in the downturn in the luminosity after the two supernova explosions at $\sim$11\Myr. Specifically, in the last simulation snapshot, the fractional luminosity is only $\sim60\%$ of its maximum luminosity, before the two supernova. Importantly, the stellar association is nearly entirely un-embedded by the time the first supernova explosion occurs. Therefore, pre-supernova feedback is efficient to de-embed most stars and remove dusty sight lines, as noted in \citet{grudic_dynamics_2022, guszejnov_effects_2022}.

We also highlight the convergence of each of the properties considered here around the 6 \Myr mark, highlighted by the gray line. At this time, each of the fractional measurements are all right around 50\%. 
While the the common thread is clearly ``the emergence of genuinely massive stars'', more detailed analysis is needed to isolate the relative importance of the luminosity, the ionizing photons, and the protostellar feedback \citep[e.g.,][]{grudic_dynamics_2022}.

\subsection{Tracking the Evolution of Formation Time and Embeddedness for Individual Stars} \label{sec:star_bed}

In the previous section, we presented a connection between the number of massive stars, and the overall extinction of the stellar association. From Figure~\ref{fig:sim_av_vs_age}, it is clear that pre-supernova feedback from massive stars is responsible for de-embedding the stars overall.
In this section, we turn to tracking the behavior of individual stars' embeddeness rather than the ensemble. Specifically, we consider the evolution of a star's typical extinction ${\widetilde{A_{\rm V}}}$ as a function of their mass and age, calculated in reference to the time when they reach their maximum mass.

\begin{figure*}[ht]
    \centering
    \includegraphics[width=0.98\textwidth]{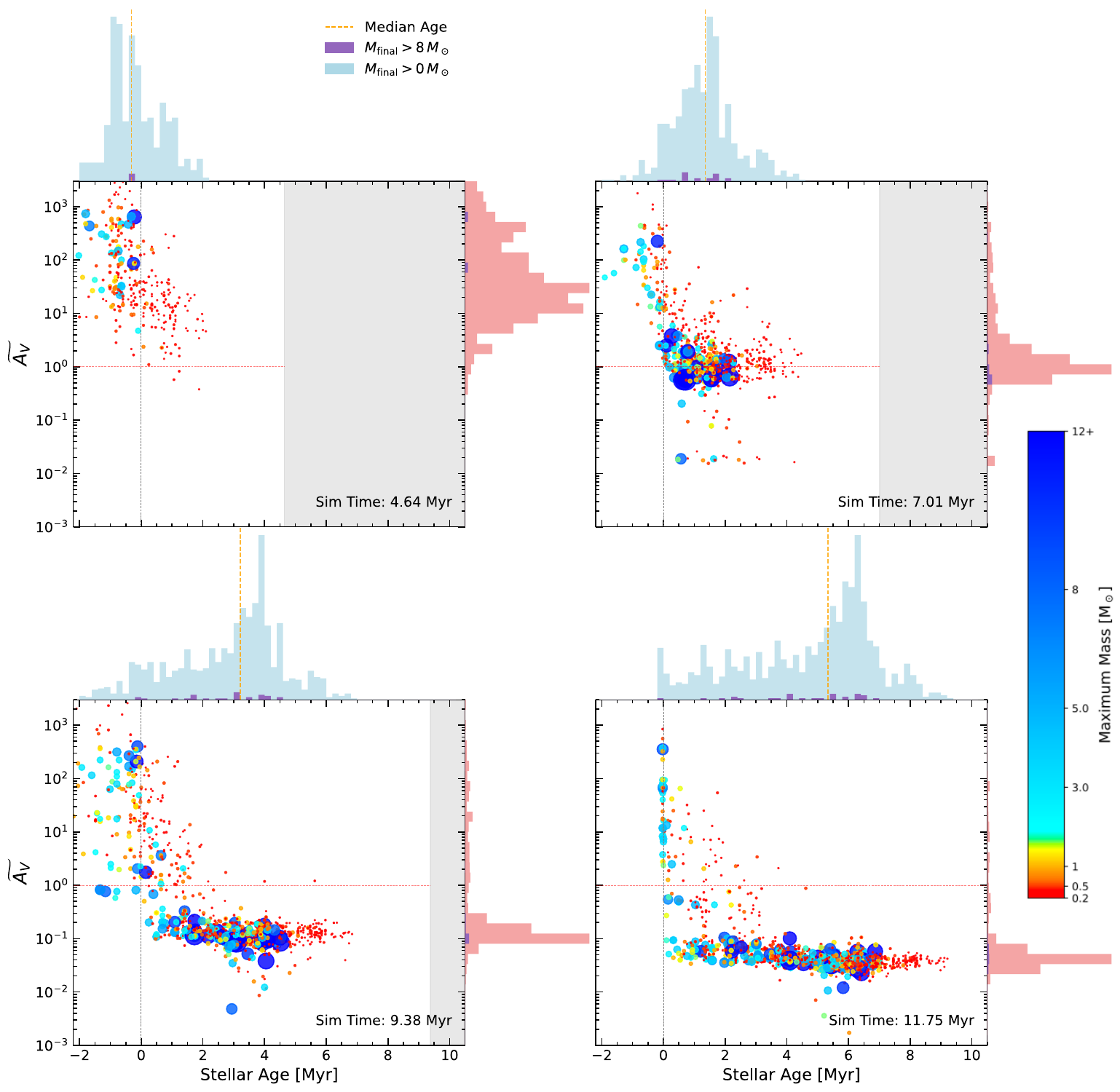}
    \caption{Each star's extinction as a function of stellar age, for four simulation snapshots. A movie showing each snapshot can be found in the online version of the manuscript. The extinction is taken to be the median value across all viewing angles, and age zero corresponds to the time the star reaches its maximum mass (Figure~\ref{fig:age_determ}) which is shown by the dashed vertical line. The horizontal red line at ${\widetilde{A_{\rm V}}} = 1$, corresponds to the red points in Figure~\ref{fig:sim_av_vs_age}. In each snapshot, the extinction decreased exponentially, with a precipitous decay around the time the star reaching it's maximum mass. Additionally, the time a star remains heavily embedded appears to depend on stellar mass, where most red points stay above the ${\widetilde{A_{\rm V}}} = 1$ line well after their zero age, compared to stars greater than 8 M$_{\odot}$ that tend to fall below this threshold right around this zero age. We discuss this trend further in Figure~\ref{fig:stellar_extinction_tracks}.} Top histograms are number counts for stars of different ages with blue being all stars and purple representing stars larger than 8 M$_{\odot}$. Side histograms are number counts of stars at a given extinction. We see star formation occurring across the entire duration of the simulation, and high mass stars form throughout. The movie can also be found at \doi{https://doi.org/10.5281/zenodo.17156032}. 
    \label{fig:av_vs_stellar_age}
\end{figure*}

We present these results in Figure~\ref{fig:av_vs_stellar_age}, which shows the value of ${\widetilde{A_{\rm V}}}$ as a function of stellar age, in four simulation snapshots; a movie for all snapshots is available as supplementary material\footnote{We also include a link to the online movie here: \href{here}{https://zenodo.org/records/17156032}}. Each star is plotted as a point that is colored and scaled in size by the star's maximum mass, using the same color bar as Figure~\ref{fig:age_determ}, such that a star keeps a consistent color and size across all plots. 
The dashed vertical line indicates zero age, when a star's host sink particle reaches maximum mass and stops accreting; stars to the left of this line (negative age) are accreting sink particles. The grey area on the right of each panel indicates times longer than the run-time of the simulation at each timestep. 

The density distribution histograms on the top and right axes track the bulk distributions in stellar age and ${\widetilde{A_{\rm V}}}$, respectively. The top plots all stars (blue), and the subset of massive stars ($>\!8\,\Msun$; purple). The vertical gold line in the histogram (and associated triangle) indicates the median age of the stars that exist at that timestep, which is typically much less than the simulation age (c.f., the median age at $t_{sim}\!=\!7\,\Myr$ is only 1.35 Myr). 
These histograms reinforce the general trend shown in Figure~\ref{fig:sim_av_vs_age} --- extended star formation and a shift towards lower extinction with increasing time.

We first focus on the overall timing of star formation throughout the simulation, best seen in the age histograms on the top of each plot. As discussed in previous STARFORGE works \citep[e.g.,][]{grudic_dynamics_2022, millstone_coevolution_2023} and in Section~\ref{sec:age_determ}, stars form across essentially the entire duration of the simulation. There is a clear peak at $t_{sim}\!\approx\!5-6\,\Myr$, but it is both preceded and followed by on-going star formation, as can best be seen in the blue histogram for the oldest timestep (bottom right, as well as Figure~\ref{fig:age_determ}). This extended star formation is consistent with persistence of residual dense gas, as was clearly visible in the snapshot images in Figure~\ref{fig:sim_av_vs_age}. 

We now turn to the joint distributions of age and extinction, as traced by individual stars. We encourage readers to view the movie version of Figure~\ref{fig:av_vs_stellar_age}, where one can track each star in higher snapshot fidelity than is possible in the 4 panel still shots. Throughout the simulation, young stars first appear in dense gas with high characteristic ${\widetilde{A_{\rm V}}}$, consistent with forming inside of dense molecular gas. Therefore, as stars form, they appear in the upper left corner of the plot. However, over time they become orders of magnitude less obscured (moving down the y-axis), typically with a notable transition near the time they stop accreting mass, at the zero-age boundary. This pattern repeats for stars formed throughout the simulation. There is also a more gradual, but steady decrease in the typical extinction of the post-accretion ($t_{age}>0$) stars, due to the on-going loss of dense dusty gas throughout the cloud.



The transition in extinction at zero age is to some degree expected, because we have defined zero age to be the moment the sink particle is no longer accreting. In many cases, this transition marks a natural inflection point. When accretion ends, local gas supply has likely diminished, and stellar feedback becomes more effective at clearing the surrounding material. As a result, extinction tends to peak around this time and then decline, although the precise timing of that decline appears to depend on the stars properties.

\begin{figure}[t]
    \centering
    \includegraphics[width=0.47\textwidth]{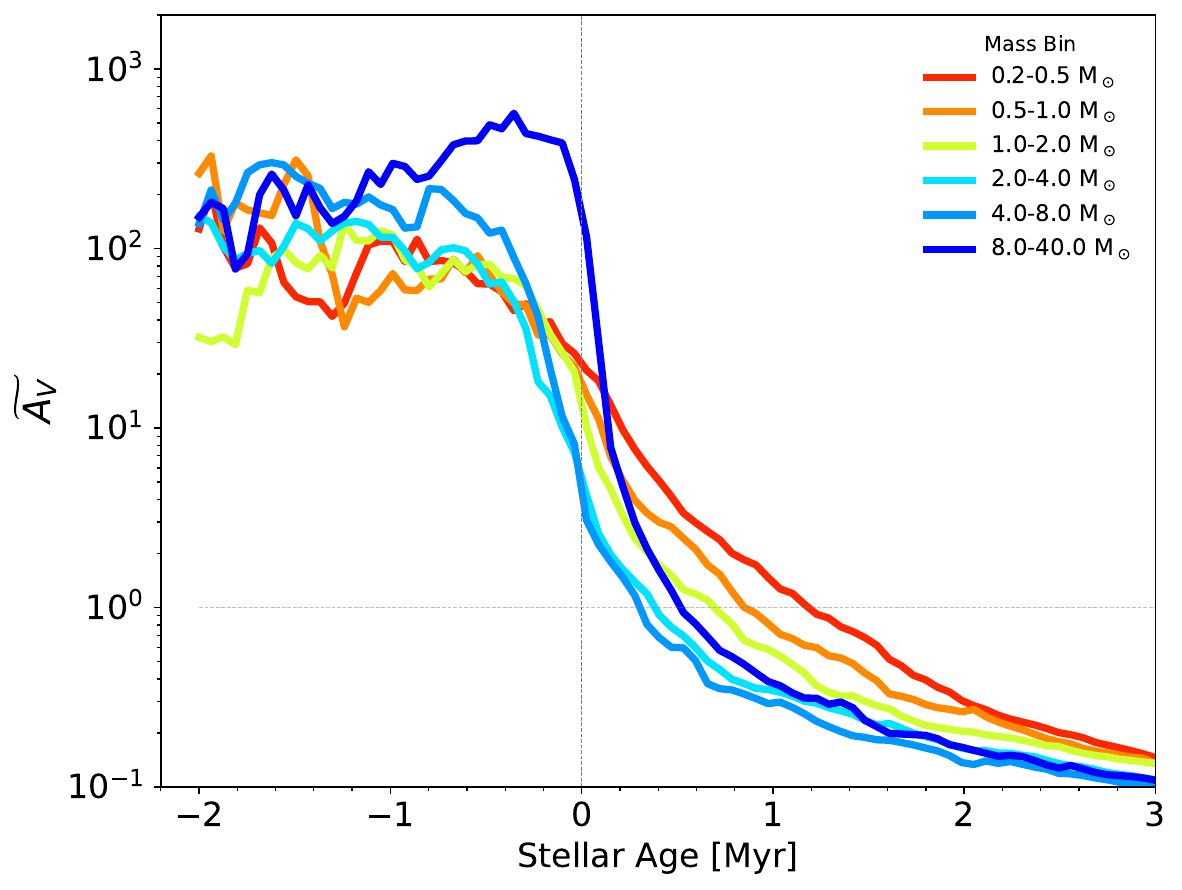}
    \caption{Stellar extinction tracks as a function of stellar age, binned by final mass. Lines represent the median values of stars in the mass bin, with colors corresponding to the final stellar mass in Figure~\ref{fig:age_determ}. There is a clear trend with when a stars begins decrease in ${\widetilde{A_{\rm V}}}$ and the final mass of the star, as well as how long it takes that star to become un-embedded. }
    \label{fig:stellar_extinction_tracks}
\end{figure}

\subsubsection{\texorpdfstring{Mass-dependence of ${\widetilde{A_{\rm V}}}(t_{\text{age}})$}{Mass-dependence of Av(t)}}

Visual inspection of Figure~\ref{fig:av_vs_stellar_age} suggests that the general correlation between the drop in a star's embedding and the end of its mass accretion may also depend on the star's final mass. There is a significant population of low mass stars (red points; $<0.5$ M$_\odot$) that remains highly embedded even after accretion stops (upper right quadrant), at all time steps. In contrast, there are essentially no higher mass stars found in the upper right quadrants, at any time step. This difference strongly suggests that low and high mass stars do not become un-obscured through identical mechanisms.

We explore this possibility more quantitatively in Figure~\ref{fig:stellar_extinction_tracks}, where we calculate the median value of ${\widetilde{A_{\rm V}}}$ as a function of stellar age for all stars in given mass bin, with colors mapping to those in Figure~\ref{fig:av_vs_stellar_age} and Figure~\ref{fig:age_determ}. These lines track the typical trajectory of stars in the panels of Figure~\ref{fig:av_vs_stellar_age}, tracing their obscuration over their lifetimes, no matter when they were formed.  

All of the trajectories in Figure~\ref{fig:stellar_extinction_tracks} show broadly similar behavior, in line of expectations from Figure~\ref{fig:av_vs_stellar_age}. Namely, stars are highly embedded for the majority of their protostellar accretion phase, and then emerge from the dusty gas. However, we also now see clear mass trends in how long stars stay heavily embedded, and how sharply their embedding drops as they reach their maximum mass. 

Regardless of their mass, stars in Figure~\ref{fig:stellar_extinction_tracks} start at similar levels of embedding. However, the low and high mass stars emerge from dusty sight-lines at different rates. High mass stars ($>8$ M$_\odot$) stay deeply embedded for longer, up until the moment they reach their maximum mass, but then become un-embedded much faster --- dropping by $\Delta\widetilde{A_{\rm V}}\sim200$ in a fraction of a Myr. In contrast, low mass stars ($<1\msun$) have more gradual evolution, de-embedding well before they reach their maximum mass, and then typically taking another $\sim1.20$ \Myr to reach ${\widetilde{A_{\rm V}} <1}$, roughly 2 times longer than that for massive stars. 

\subsubsection{\texorpdfstring{What is Driving the Mass-dependence of $\widetilde{A_{\rm V}}(t_{\text{age}})$?}{What is Driving the Mass-dependence of Av(t)?}}

There are three general classes of mechanisms that might drive stars' emergence from deep obscuration: (1) global feedback; (2) local feedback; and (3) drift. In the first, stars may become systematically less extincted because of the global loss of dusty gas from the cluster, due to the cumulative effects of feedback. In the second ``local'' scenario, most of the embedding is actually from dust close to an individual star, where it can be cleared out by a star's own feedback. In the final ``drift'' scenario, some of the emergence from embeddedness may be due to stars and gas clumps moving apart, because of velocity differences and/or the changing structure of the turbulent gas. In this scenario, even modest relative velocities can move a star out of a dense gas cloud in a short time, provided the cloud is small.  

The plots in Figures~\ref{fig:av_vs_stellar_age}~\&~\ref{fig:stellar_extinction_tracks} strongly suggest that for massive stars, emergence from embedding is driven by their own feedback into their local natal gas. Almost every massive star (i.e, dark blue) plummets in ${\widetilde{A_{\rm V}}}$ almost exactly when they reach their maximum mass, and reach ${\widetilde{A_{\rm V}} <1}$ by a median 0.6 \Myr after the maximum mass is reached. This behavior remains consistent across time and is independent of the extinction of the overall association, strongly suggesting that the transition is driven by local, rather than global, feedback. Dense gas is needed for the massive star to form, but the removal of that gas is largely controlled by the star itself.  

Not only do the massive stars drive their own emergence from the dusty gas, they also appear to control their early evolution as well. Unlike the majority of stars in the simulation, the most massive stars' obscuration initially \textit{increases} with time, presumably due to gravitational contraction of the surrounding gas cloud. This contraction provides fuel for the massive star's on-going gas accretion (left panel of Figure~\ref{fig:mass_accretion_vs_time}), up until the star's own feedback shuts off the accretion and prevents any further increase in the star's mass. The fact that the embedding plummets at exactly the same time strongly suggests that, for these massive stars at least, the majority of their line-of-sight extinction is extremely local.

The behavior of the low-mass stars is quite different, however. These stars will have limited ability to drive stellar feedback, given that the available energy during the pre-main sequence phase is strongly mass dependent (Section~\ref{sec:Drivers}), and their final luminosities are low. It therefore is unlikely that their reduction in embeddedness is feedback driven. 

However, low-mass stars are not entirely passive in shaping their local environments. Even though their radiative output is minimal, these stars can drive powerful collimated outflows during the protostellar phase \citep[e.g.,][]{frank_jets_2014}, which have been shown to efficiently expel nearby gas on small scales \citep[e.g.,][]{offner_impact_2017}. Because the outflow momentum and radiative output are directly to the stellar mass \citep{offner_impact_2017}, low-mass protostars inject substantially less momentum per unit time than their high-mass counterparts, meaning that any reduction in extinction proceeds more slowly and may not reach large enough scales to fully de-embed the star. Indeed, in Figure~\ref{fig:stellar_extinction_tracks}, we see the evolution of extinction for low-mass stars takes place over a longer timescale, with no strong features in $\widetilde{A_V}$ associated with reaching their maximum mass. 

The evolution of embeddedness for low-mass stars is therefore shaped by a combination of this limited local feedback, and external environmental factors. While outflows may partially clear nearby gas, these low-mass stars lack the ability to regulate their broader surroundings. Instead, their emergence can then reflect either drift relative to dense filaments, changes in the gas structure driven by turbulence, or the cumulative influence of more massive neighbors. Interestingly, this empirically leads to a situation where it is the lower mass stars that remain embedded for longer, well-past the time that the massive stars have become unobscured; this is in contrast to scenarios where the formation of massive stars clears out the obscuration for all stars.

While we have stressed the strong impact of local embeddeness and feedback for massive stars, there are global effects as well. The distributions of individual stars in Figure~\ref{fig:av_vs_stellar_age}, along with the red histograms of ${\widetilde{A_{\rm V}}}$ on the right axes, show that after $\sim$6\Myr there is a population of weakly-embedded stars that have reached their maximum mass, and that all have similar median embeddedness. These are stars that are currently outside of any dense gas clump (and thus no longer accreting mass), whose line-of-sight extinctions are now set by the overall extinction properties of the global gas distribution. The median extinction of this population decreases steadily with time, due to the global evolution of gas in the system due to on-going feedback. As such, in Figure~\ref{fig:stellar_extinction_tracks} the late time evolution of ${\widetilde{A_{\rm V}}}$ for most stars looks quite similar, particularly for the more massive stars that have eliminated any local sources of obscuration, and are now subject only to the statistical properties of extinction in the cluster as a whole. 

In summary, the change of embeddedness is local to the high mass stars. For massive stars, the energy required to overcome the accretion is larger due to larger gravitational potential of the larger system. Once accretion stops, the large radiation pressure then helps to clear out the region around the star. In contrast, the outflows from low-mass stars take place on longer time-scales, and the stars are free to move relative to the gas, and can therefore de-embed in part through geometry, regardless of the overall gas reservoir in the system.

\subsection{How is the Time Evolution of Embdeddedness Impacted by Cloud Mass?} \label{sec:sob_ts}

There are a number of timescales we expect to operate during cluster formation, including the free-fall time $\tau_{ff}$ for gravitational collapse, the gas accretion timescale for protostars, and the stellar evolution timescales associated with feedback (i.e., the emergence of jets, ionizing radiation, and supernovae). The latter of these timescales are connected closely to the properties of individual stars, and should be relatively similar from cluster to cluster, at fixed metallicity. In contrast, the free-fall and gas accretion times are both dependent on the initial properties of the molecular cloud as a whole, including its mass. 

Here, we investigate the empirical mass-dependence of the timescale for cluster embedding. As discussed in Section~\ref{sec:clst_bed}, the fiducial STARFORGE simulation is based on a $2 \times 10^4$ M$_{\odot}$ cloud, $10\,\pc$ in radius. The free-fall time depends on the initial cloud density $\rho_0$ as $\tau_{ff}=(3\pi/32{\rm G}\rho_0)^{1/2}$, which for the fiducial model corresponds to 3.7 \Myr.
We now compare the timescales of the fiducial simulation to a second, more massive cloud simulation beginning with $2 \times 10^5$ M$_{\odot}$ and a $30\,\pc$ radius. The initial density of this higher mass simulation is a factor of 2.7$\times$ smaller than the low mass simulation, leading to a 1.6$\times$ longer free-fall time for the massive cloud. Both simulations have identical initial virial parameters, such that their crossing times at the characteristic velocity dispersions scale linearly with the free-fall time. We therefore expect that the more massive cloud should take longer to collapse, and should have gas available for accretion over a longer timescale.

\begin{figure*}[ht]
    \centering
    \includegraphics[width=0.98\textwidth]{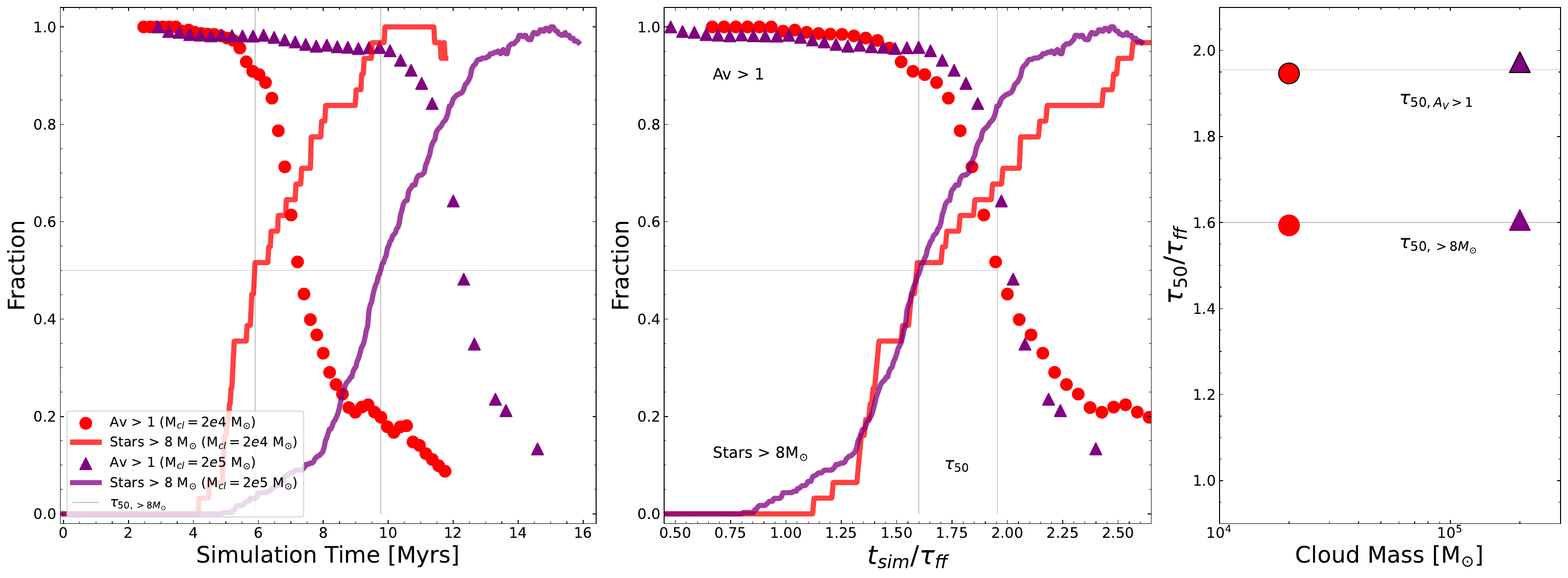}
    \caption{Comparison between two STARFORGE simulations, differing in molecular cloud mass. The fiducial $2 \times 10^4$ M$_{\odot}$ cloud is shown in red, while the $2 \times 10^4$ M$_{\odot}$ cloud is shown in purple. The left panel correspond to Figure~\ref{fig:sim_av_vs_age}, with points showing the fraction of stars with ${\widetilde{A_{\rm V}}}$ greater than 1, while the solid lines show the fraction of stars greater than 8 M$_{\odot}$ which have begun their proto-stellar stage. The middle panel shows the same as the left, but is normalized by the respective cloud free-fall times, $\tau_{ff}$, demonstrating that the start of emergence is related to initial cloud mass and density. The vertical gray lines correspond to $\tau_{50, >8M_{\odot}}$ and $\tau_{50, A_{\rm V} > 1}$, which denote the time where $50\%$ of of the stars greater than 8 M$_{\odot}$ have begun their proto-stellar stages, and the time when $50\%$ of of the stars have a ${\widetilde{A_{\rm V}}}$ greater than one. These metrics are also shown in the right panel. In both simulations, the extinction is tied to the point in the star-forming event where the high-mass stars form.  }
    \label{fig:normed_dsims}
\end{figure*}

We test this assumption in Figure~\ref{fig:normed_dsims} by generating plots similar to the bottom of Figure~\ref{fig:sim_av_vs_age} for both the low mass (red) and high mass (purple) simulations. The filled points indicate the time evolution of the fraction of stars with ${\widetilde{A_{\rm V}}}\!>\!1$, which for both simulations starts at one (all stars highly obscured) and declines to small values (little net extinction) by the end of the simulation. The solid lines indicate the fraction of high mass ($>\!8\,\msun$) stars $f_{>8\msun}$ at each timestep (same accounting as right hand panel in Figure~\ref{fig:fid_counting}\footnote{We note that we could have chosen to measure the fraction of high mass stars which have reached their maximum mass, consistent with the age zero from Section~\ref{sec:age_determ}. If we had made this distinction, $\tau_{50, >8M_{\odot}}$ in the Fiducial model increases slightly to 1.72 $\tau_{ff}$.}) as a function of time, which increases from zero to one for both the low mass (red) and high mass (purple) cluster simulations, and decreases as stars explode and undergo supernovae.

The left-hand panel of Figure~\ref{fig:normed_dsims} shows the evolution of the embedded and the high mass star populations in raw simulation time. As expected from the initial free-fall time calculations, the characteristic formation timescale of the high mass cluster is significantly longer. Massive stars take longer to form, and all stars take longer to emerge from the dust.

The fact that all the curves in Figure~\ref{fig:normed_dsims} have similar shapes suggests that they may scale with the free-fall time. We test this possibility in the center panel, where we rescale the simulation time by $\tau_{ff}$. We find that the bulk of the evolution in $f({\widetilde{A_{\rm V}}}\!>\!1)$ and $f(>8\msun)$ is quite similar, once the scaling in the free-fall time is removed. We note that we do not expect the trends for the two clouds to be in perfect agreement, given that there will be second-order effects from specifics in the cloud substructure, and that some feedback timescales such as ''time to first supernova after a massive star forms'' are not going to be affected by the overall cluster dynamics.

We quantify the scaling with free-fall time by calculating the times at which the fractions $f({\widetilde{A_{\rm V}}}\!>\!1)$ and $f(>8\msun)$ reach 50\%, as indicated by the thin grey lines in the left and center panels. We then plot these two quantities as function of cloud mass in the right most panel of Figure~\ref{fig:normed_dsims}. When considering the scaling of the cloud free fall time, we find that the point that 50\% of the massive stars form ($\tau_{50, >8M_{\odot}}$) occurs at 1.6 $\tau_{ff}$ in both clouds. 

In addition to $\tau_{50, >8M_{\odot}}$ being consistent between the two clouds, we also find the point where $50\%$ of stars have an $\widetilde{A_{\rm V}}$ greater than 1 ($\tau_{50, A_{\rm V} > 1}$), to also scale with the cloud free fall time. We measure $\tau_{50, A_{\rm V} > 1}$ to be 1.97 $\tau_{ff}$ for the $2 \times 10^5 \rm{M}_{\odot}$ cloud and 1.94 $\tau_{ff}$ in the $2 \times 10^4 \rm{M}_{\odot}$ cloud. Given that the $\tau_{50, >8M_{\odot}}$ in both clouds is 1.6 $\tau_{ff}$, this slight difference in $\tau_{50, A_{\rm V} > 1}$ between the two simulations means that the emergence occurs slightly faster in the lower mass cloud than in the high mass cloud. 

We note that the exact scaling of 1.6 free-fall times is likely to change with different initial conditions, or different chemical prescriptions. While the exact scaling may change, we expect the overall correlation to hold. Physically, we demonstrate the onset of massive star formation, and the subsequent feedback associated with them, to be the driving mechanism of gas clearance. While the formation time of massive stars is likely dependent on the initial conditions of the simulation, given the spread in stellar age seen here, the length of this formation time should still scale with the free fall time of the cloud. 

This result once again re-enforces that de-embedding is governed by the local impact of massive stars. Once a massive star forms, their feedback -- ionization, radiation pressure, and winds -- rapidly evacuates its immediate surroundings, regardless of the mean cloud density. In this regime, the de-embedding timescale reflects the time required to form the massive stars themselves, not the efficiency of subsequent large-scale feedback. For the two cloud masses explored here, this local mechanism dominates, yielding a nearly self-similar emergence time in units of $\tau_{\mathrm{ff}}$. However, more analysis is needed for other simulations with different cloud masses and different initial conditions. At significantly higher densities or metallicities, we expect this balance may shift, but within our parameter space, the onset of de-embedding is tightly linked to the timing of massive star formation.

\section{Observational Measures of Embeddedness}\label{sec:observations}

Throughout this work, we have focused on measures of embeddedness from the lens where we know the intrinsic properties of all aspects of the simulation. This framework has allowed us to probe the relationships between newly formed stars, their feedback, and the timescales of embeddedness. However, these ground truths are not a luxury afforded to those in the observational community. We therefore now approach the simulation from another angle, re-framing our analysis to mirror the observational approach. 

In the previous section, this framework led us to conclude that individual stars transition from deeply embedded to exposed very rapidly once they reach their maximum mass, and that this transition is driven primarily by local feedback from massive stars rather than by global gas removal (Section~\ref{sec:results}). To connect this physical picture to what observers actually measure, we must determine whether such locally driven emergence would be imprinted in the diagnostics used to infer embedded lifetimes. We therefore model the standard tracers used to infer embedded timescales and star–formation rates from our simulation, and measure how long each tracer is dominant, peaks, and fades relative to the physically defined de–embedding of the stars presented in the previous section. By comparing these synthetic timescales to those inferred from real systems, we can test both whether the short, local emergence phase around massive stars leaves a clear imprint on the luminosity evolution and whether the STARFORGE model is capturing the same physics that governs observed embedded clusters.

While trying to observe star formation, there are specific times throughout the process where different tracers dominate the emission from the region \citep[e.g.,][]{kruijssen_uncertainty_2018, kim_duration_2021}. These different tracers can then lead to different measured embedded timescales, depending on the how long each tracer is viable. In this section, we discuss three main observational tracers often used: 1) IR emission from irradiated dust, 2) H$\alpha$ emission from hydrogen re-combination, and 3) UV/Optical emission from stars. We will then compare the timescales for which different wavelength regimes are observable. 

\subsection{Infrared Emission} \label{sec:ir}

When stars are obscured, their light is absorbed by the surrounding dust. The warmed dust then re-emits this energy as thermal emission at mid- and far-infrared (MIR/FIR) wavelengths. This dust emission therefore has become widely used as a way to find young, but shrouded, stellar clusters early in their evolution. 
In this section, we calculate this emission from warm dust, and the fraction of reprocessed stellar luminosity as a function of time, with the goal of measuring the timescales over which MIR/FIR emission might be detected for young clusters. 

To model the emergent radiation from the dust in our simulation, we perform post-processing radiative transfer calculations using the \texttt{SKIRT} code \citep{baes_efficient_2011, camps_skirt_2015, camps_skirt_2020}. \texttt{SKIRT} is a three-dimensional Monte-Carlo radiative transfer code that tracks the propagation, absorption, scattering, and re-emission of photon packets through arbitrary dusty geometries. \texttt{SKIRT} includes both stellar and background sources, and computes self-consistent dust heating and emission assuming local thermodynamic equilibrium and stochastic heating \citep{baes_skirt_2015}.

The primary radiation is provided by both stellar particles and a diffuse interstellar radiation field (ISRF). Stellar sources are imported from STARFORGE outputs as individual particles, each with a position, velocity, mass, age, and metallicity. We then assign each star an SED drawn from the \citet{bruzual_stellar_2003} simple stellar population models. 
In addition to the stars, we include a spherical background radiation field that approximates the ISRF, following the multi-component modified blackbody model from \citet{zucconi_dust_2001} as updated by \citet{bate_combining_2015}. 

The absorbing and emitting medium consists of the gas cells in the simulation, which are imported into \texttt{SKIRT}. Each cell provides a position, mass, metallicity, and dust temperature, where \texttt{SKIRT} then internally computes the dust mass assuming a fixed dust-to-metals ratio of 0.4. We adopt the \citet{draine_infrared_2007} dust mixture, and the dust is discretized onto a Voronoi mesh with 500,000 sites, where photon packets are propagated through the grid. For accuracy, we include dust heating by the CMB, though it has a negligible effect. For emission, we adopt a logarithmically spaced wavelength grid from 0.1 to 2000\micron with 200 bins, ensuring accurate coverage of both stellar and IR re-emission peaks. 

The dust luminosity is calculated internally by \texttt{SKIRT} as the total bolometric energy absorbed, and subsequently re-emitted by dust grains. Specifically, \texttt{SKIRT} determines the absorbed luminosity in each spatial cell $m$ by integrating the product of the radiation field, absorption opacity, and path length over all wavelength bins $\ell$:
\begin{equation}
    L^\mathrm{abs}_{\mathrm{bol},m} = \sum_\ell k^\mathrm{abs}_{\ell,m} \left(L \Delta s\right)_{\ell,m} ,
\end{equation}
where $k^\mathrm{abs}_{\ell,m}$ is the absorption opacity in cell $m$ at wavelength bin $\ell$, and $\left(L \Delta s\right)_{\ell,m}$ is the path-integrated packet luminosity accumulated in that bin. The total dust luminosity is then obtained by summing over all cells.

While \texttt{SKIRT} includes the capability of including separate contributions from primary and secondary emission cycles \citep{camps_skirt_2015}, in this work we choose to focus on the dust luminosity resulting from the absorption of light from stars and the ISRF, and all dust luminosities presented are those of the primary emission cycles. Therefore, while the luminosity values presented here represent a lower limit, the overall shape and timescales of the dust luminosity, are representative of the total infrared emission. 

Based on the absorbed energy, dust properties, and the local radiation field, \texttt{skirt}
calculates the dust emissivity spectrum $j_\nu(\vec{r})$ in each cell. Photon packets are then emitted according to this emissivity distribution. The total emergent dust luminosity is given by:
\begin{equation} \label{eq:L_dust}
    L_{\rm dust} = \int_V \int_\nu j_\nu(\vec{r})\,\mathrm{d}\nu\,\mathrm{d}V ,
\end{equation}
where $j_\nu(\vec{r}) = \kappa_\nu \rho(\vec{r}) B_\nu(T_d(\vec{r}))$ for grains in thermal equilibrium, or is derived from the full temperature probability distribution for stochastically heated grains \citep{baes_efficient_2011, camps_skirt_2015, verstocken_skirt_2017}. The resulting $L_{\rm dust}$ includes the integrated emission from all dust species and temperatures present in the simulation volume.

\begin{figure*}
    \centering
    \includegraphics[width=0.98\textwidth]{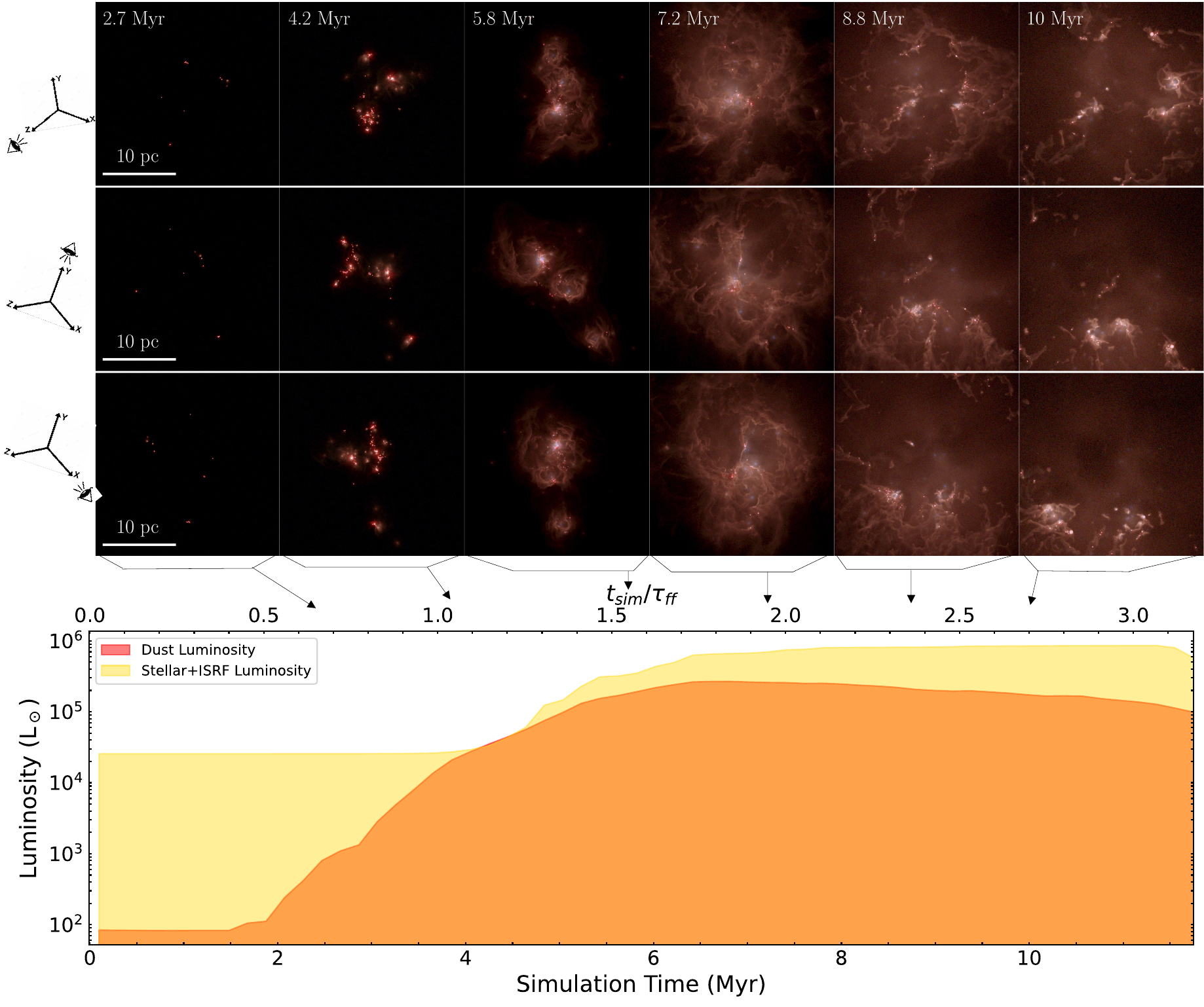}
    \caption{Top: three-color (R: 80-120 \micron, G: 6-13 \micron\ B: 20-30 \micron) composite images of the STARFORGE simulation viewed from 3 different angles at 500 \pc with time increasing left to right. Bottom: time evolution of the total dust luminosity in solar lumins (red) compared to the luminosity from stars and the ISRF (yellow). We include a larger, high-resolution image of one viewing angle from the 7.2 \Myr snapshot in Appendix~\ref{appendix}.}
    \label{fig:dust_lum_vs_time}
\end{figure*}

\subsubsection{Total Dust Luminosity}
The time evolution of the total dust luminosity across our simulation snapshots is shown in bottom panel of Figure~\ref{fig:dust_lum_vs_time}. Shown in red is the total luminosity, while yellow shows the combined luminosity of the ISFR and stars. At the start of the simulation when stars have yet to form, the energy budget is dominated by the ISRF, and the total dust luminosity is $\sim80\rm{L}_\odot$. This initial dust luminosity remains relatively constant until the first stars begin to form and heat the dust. The dust luminosity then increases gradually as more stars form, and peaks at 6.8 Myrs (1.68 $\tau_{ff}$) with the formation of large numbers of massive stars (see upper right panel of Figure~\ref{fig:av_vs_stellar_age}). 

For visualization purposes, we also generate synthetic images of dust luminosity calculated by \texttt{skirt} in the STARFORGE simulation, which are shown in the top panel of Figure~\ref{fig:dust_lum_vs_time}. These images are generated from the same simulation snapshots as in Figure~\ref{fig:sim_av_vs_age}, with time increasing from left to right, and each row presenting the same simulation output viewed at a different orientation. Each image spans 25 pc in the x and y direction, covered by 25 Megapixels. For these images, we set a distance of 500 \pc to the source, giving a angular resolution of $\sim2$ arcseconds per pixel, roughly consistent with JWSTs instruments \citep[][]{rieke_mid-infrared_2015, wright_mid-infrared_2023}. The three-color composite images- R: 80-120 \micron, G: 6-13 \micron B: 20-30 \micron- were chosen to best represent the SED, where red is the peak of the SED showing cold dust, green shows the primary PAH features, and blue shows the features arising from hot dust. We also include one larger, high-resolution image of the 7.2 \Myr snapshot in Appendix~\ref{appendix} to best showcase the filamentary dust structure that emerges in STARFORGE. 

Images were created using \texttt{ds9} \citep{saoimage_ds9}. Each channel was scaled such that they are equally-weighted by integrated luminosity in the snapshot, and to compensate for the evolving total dust luminosity across snapshots (see Figure \ref{fig:dust_lum_vs_time}, bottom panel; i.e. 2360 has the highest maximum scaling, while 1080 has the lowest). Each channel was logarithmically-scaled, with a varying log-exponent (ranging from 3000-10,000), depending on the dynamic range of the image, such that a higher exponent was used for snapshots with a large amount of fine structure at low-surface brightness. 

In these composite images, the different colors reveal distinct physical processes and evolutionary stages within the dusty star-forming region. The bright red knots, which dominate early in the evolution, represent dense, cold dust concentrated around young protostars still deeply embedded in their natal gas. These compact structures correspond to the regions of highest extinction, and their prominence reflects ongoing accretion and gravitational collapse. In contrast, the blue regions trace hot dust that has been heated by the intense radiation of recently formed massive stars. This emission, arising primarily in the 20–30 \micron ~ range, is characteristic of photodissociation regions (PDRs) where stellar UV radiation heats and erodes surrounding material. As the system evolves, these blue features become more spatially extended, marking the emergence of heated cavities and expanding shells. The green channel highlights mid-infrared emission from PAHs, which are also clustered around regions of recent massive star formation. Together, these evolving color patterns offer a visual narrative of how feedback transforms the structure of the cloud, progressing from dense, cold star-forming cores to dispersed, UV-irradiated PDRs.

This evolving color structure reflects not only changes in local conditions but also tracks the global evolution of dust luminosity over time. As more stars form and begin heating their surroundings, the total dust luminosity increases steadily, peaking around 6.8 Myr (1.84$\tau_{\rm ff}$), coinciding with the appearance of large numbers of massive stars (see Figure~\ref{fig:av_vs_stellar_age}). This trend closely parallels the decline in median extinction shown in Figure~\ref{fig:sim_av_vs_age}, underscoring how star formation and dust heating go hand in hand during the embedded phase.

Beyond the overall rise and fall in brightness, the morphology of the dust emission undergoes a dramatic transformation. Early on, emission is confined to compact, embedded clumps, but by 10 \Myr, the gas has largely been dispersed and the stars themselves have begun to drift apart. The remaining dust appears far more diffuse, but continues to emit due to localized heating from residual star-forming activity. These late-time clumps stand out clearly in the upper panels of Figure~\ref{fig:dust_lum_vs_time}, marking the final, spatially fragmented stages of clustered star formation.

\subsubsection{\texorpdfstring{24$\micron$ Emission}{24 micron Emission}}

In addition to the bolometric dust luminosity presented above, we also choose to highlight $24\micron$ emission, which has become widely used as a tracer of embedded star formation since the launch of \textit{Spitzer} \citep[e.g.,][]{schinnerer_molecular_2024}. This mid-infrared band is dominated by emission from warm, large dust particles, with little emission from more fragile polyaromatic hydrocarbons (PAHs) \citep{whitcomb_star_2023} making the 24$\micron$ emission easier to interpret. Scattering is also negligible at $24\micron$, with an albedo of only $4.6\times 10^{-4}$, making Eq. \ref{eq:L_dust} a good approximation. Additionally, the F2100W filter on JWST serves a similar purpose, and tends to track the same physical processes \citep{ramambason_duration_2025}.

Using the same framework used to make the images, and calculate the dust luminosity in Section~\ref{sec:ir}, we make mock 24$\micron$ observations of the simulation using \texttt{skirt}. Here, we adopt the \textit{Spitzer} 24$\micron$ pass-band which ranges from 20.8 to 26.1 microns, integrating the dust SED only across these wavelengths to determine the 24$\micron$ luminosity, as observed by \textit{Spitzer}.

\begin{figure*}[ht]
    \centering
    \includegraphics[width=0.94\textwidth]{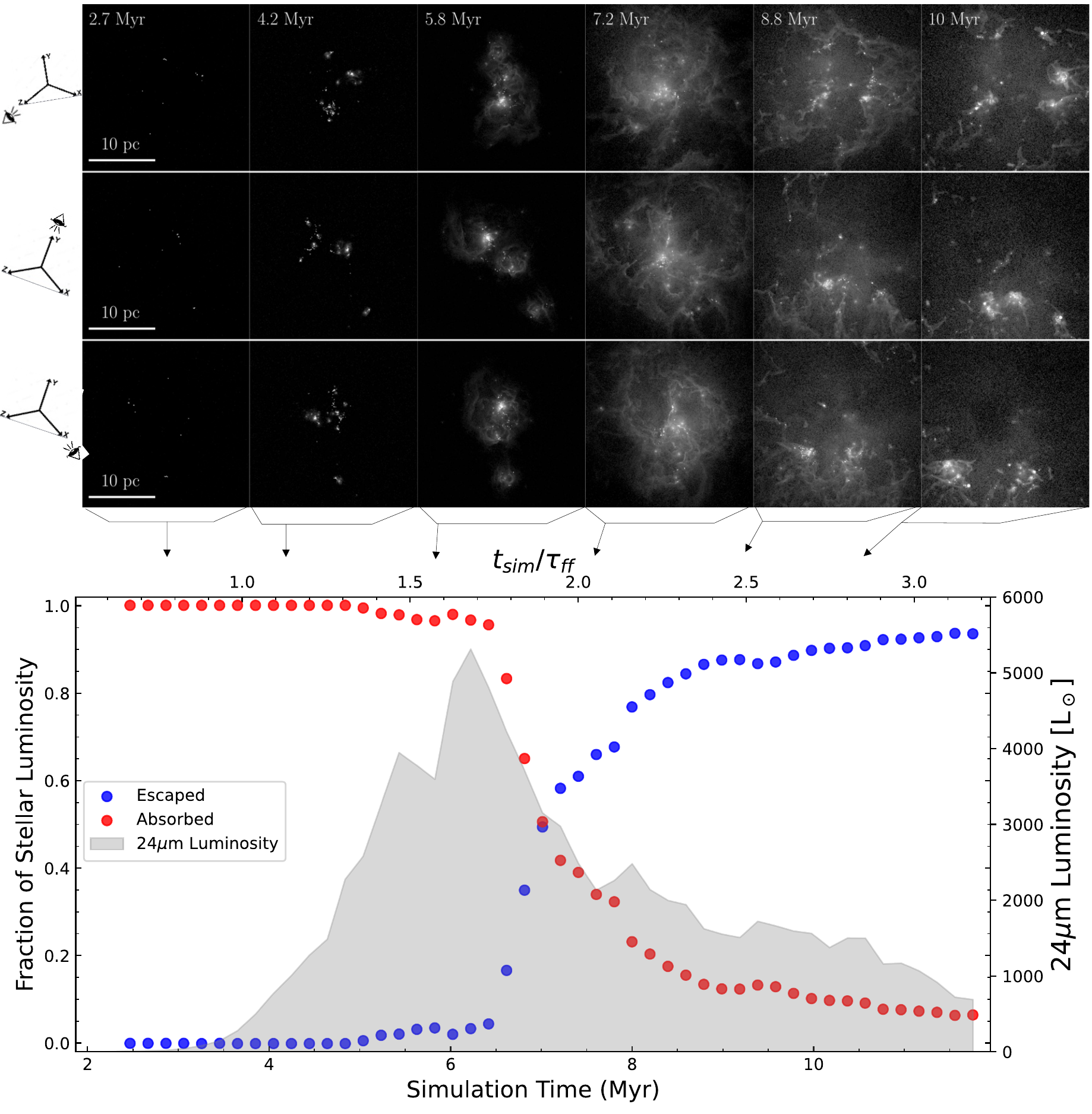}
    \caption{\textit{Top:} $24\micron$ luminosity from the \textit{Spitzer} band-pass, visualized at the same viewing angles and time steps as Figure~\ref{fig:sim_av_vs_age}. \textit{Bottom:} Fraction of stellar luminosity absorbed (red), and escaped (blue). On the secondary axis and shown in gray is the total $24\micron$ luminosity of the region as a function of time. The $24\micron$ luminosity from dust emission peaks before $20\%$ of the stellar luminosity escapes and is visible. }
    \label{fig:tf_ir}
\end{figure*}

In the top panels of Figure~\ref{fig:tf_ir}, we show a time sequence of 24$\micron$ images, generated from the same simulation snapshots as in Figure~\ref{fig:sim_av_vs_age}, with time increasing from left to right, and each row presenting the same simulation output viewed at a different orientation. The gray shaded region in the bottom panel is the integrated $24\micron$ luminosity of the region as a function of time, while the red and blue points show the fraction of the emitted V-band stellar luminosity that is absorbed by the dust and how much escapes, respectively. The absorbed fraction is derived from the fractional, cumulative stellar $A_{\rm V_{\theta,\phi}}$, integrated over all angles, where the amount escaped is simply the inverse of how much is absorbed. 

Overall, the evolution in Figure~\ref{fig:tf_ir} is quite similar to that seen for the overall dust luminosity. However, while the $24\micron$ luminosity increases gradually as more stars form just like the overall dust luminosity, the $24\micron$ luminosity peaks earlier, at 6.2 \Myr (1.68 $\tau_{ff}$) compared to the peak at 6.8 \Myr seen in Section~\ref{sec:ir}. Additionally, in contrast to the relatively smooth decline in luminosity seen in in Figure~\ref{fig:dust_lum_vs_time}, the decline in $24\micron$ luminosity from the peak is much more pronounced. In fact, by 6.8 \Myr when the total dust luminosity is peaking, the $24\micron$ luminosity is already at 70\% of its maximum. 

When the $24\micron$ begins to decrease, in the bottom panel of Figure~\ref{fig:tf_ir} we can see that over $95\%$ of the optical stellar luminosity is being absorbed, warming the dust and powering the $24\micron$ emission. However, during the period of time between $\sim$6 and $8\,\Myr$ the fraction of significantly extincted stars is steadily dropping (bottom panel of Figure~\ref{fig:sim_av_vs_age}), and the $24\micron$ luminosity drops as the star light becomes less coupled to the dust and a larger fraction of the light escapes (blue points in Figure~\ref{fig:tf_ir}). 
This low level 24$\micron$ emission seen here after the peak in star formation is consistent with that seen in \citet{kim_duration_2021}, who measure diffuse emission for up to 9 \Myr after star formation ceases. 

In total, the upper and lower panels of Figure~\ref{fig:tf_ir} confirm that compact, high-intensity 24$\micron$ emission is a signature of the embedded phase of star formation. Within STARFORGE, the 24$\micron$ luminosity is high for only a brief window of $\sim$2-3$\,\Myr$, corresponding to the period where the emission is truly embedded. There is also a fainter, but extended tail of 24$\micron$ emission that persists for an additional $\sim$5-6$\,\Myr$. However, during this long tail, the existence of 24$\micron$ emission alone does not necessarily imply the association is in the embedded phase, particularly when there is not enough spatial resolution to discern the morphology of the emission.

\subsection{\texorpdfstring{Gas Density and H$\alpha$ Emission}{Gas Density and Halpha Emission}}
\label{sec:h_alpha}

Later in their evolution, young clusters can be identified through their H$\alpha$ emission at 656~nm \citep[e.g.,][]{rydgren_observations_1979,kennicutt_star_1998}, which arises from recombination of hydrogen (H\textsc{ii}) that has been ionized by young massive stars or shocks \citep[e.g.,][]{stromgren_physical_1939,kennicutt_survey_1983,kennicutt_star_1998, tacchella_h_2022, schinnerer_molecular_2024}. H$\alpha$ emission is therefore a direct indicator of the presence of short-lived massive stars, and thus offers constraints on the age of a stellar cluster. Its morphology is also a useful indicator of the evolutionary state of gas clouds, which can range from ``tracing a dense reservoir of ionized gas'' to ``tracing recent photodissociation along extended surfaces of a disrupting turbulent molecular cloud''\citep[e.g.,][]{whitmore_using_2011,hannon_h_2022}.

Here, we calculate the H$\alpha$ luminosity and morphology for the evolving stellar association. For each gas cell, we derive the H$\alpha$ luminosity from the number density of protons ($n_p$) and electrons ($n_e$), using the recombination rate $\alpha$ such that,  

\begin{equation} \label{eq:h_alpha}
    L_{\text{H}\alpha} = 2.86 \alpha^{\text{eff}}_{\text{H}\beta} \, n_p \, n_e \, \frac{hc}{\lambda_{\text{H}\alpha}} \cdot \frac{m}{\rho},
\end{equation}
where $\alpha^{\text{eff}}_{\text{H}\beta}$ is the effective recombination coefficient for H$\beta$ emission, scaled by the H$\alpha$/H$\beta$ line ratio of 2.86, at 10000 K, taken from \citet{osterbrock_astrophysics_2006}. 

The proton number density $n_p$ is given by the product of the ionized hydrogen fraction $x_{\text{H\,\textsc{ii}}}$ and the total hydrogen number density $n_{\text{H}}$, while the electron number density $n_e$ is computed in STARFORGE. The hydrogen number density itself is derived from the gas density $\rho$ using $n_{\text{H}} = \rho / (\mu_{\text{H}} m_p)$, where $m_p$ is the proton mass and $\mu_{\text{H}}$ is the hydrogen mass fraction, measured for each gas particle based on metallically.

The energy emitted in H$\alpha$ is then computed as $hc / \lambda_{\text{H}\alpha}$, where $h$ is Planck’s constant, $c$ is the speed of light, and $\lambda_{\text{H}\alpha} = 656\,\mathrm{nm}$ is the rest-frame wavelength of the H$\alpha$ line. The final factor $m / \rho$ represents the physical volume associated with each gas particle, effectively integrating the local emissivity over that volume. The result is the total H$\alpha$ luminosity per gas particle, expressed in units of $\mathrm{erg\,s^{-1}}$.

\begin{figure*}[ht]
    \centering
    \includegraphics[width=0.97\textwidth]{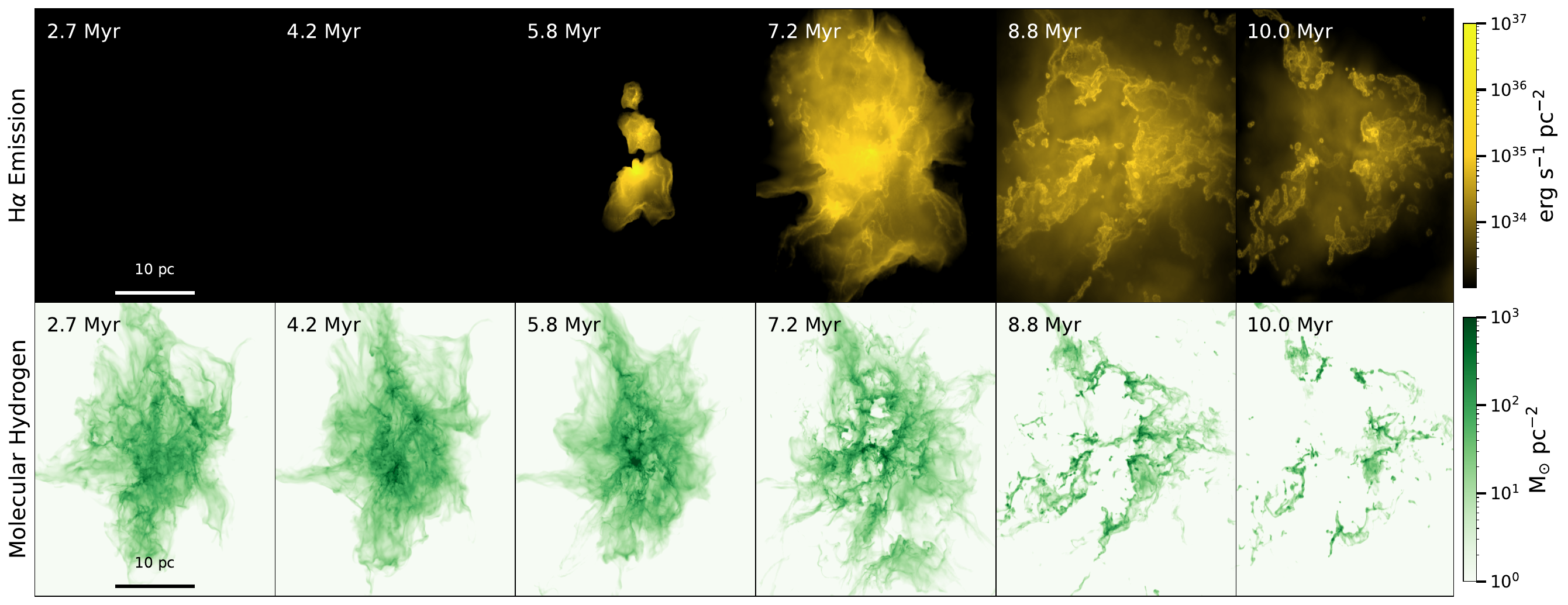}
    \caption{Surface brightness of H$_\alpha$ emission (yellow color map; top) and the surface density of molecular hydrogen (bottom; green), as a function of simulation time. The densest regions of the molecular gas are the locations of star formation, which is traced by the H$\alpha$ in the top panel. The H$\alpha$ emission also mirrors that seen in the $24\micron$ in Figure~\ref{fig:tf_ir}. }
    \label{fig:h_alpha}
\end{figure*}

Figure~\ref{fig:h_alpha} shows the resulting maps of H$\alpha$ emission at the same time steps and orientations as the top row in Figure~\ref{fig:sim_av_vs_age}. The colorbar represents the surface brightness of the emission, where yellow corresponds to brighter regions. To provide context, we also show maps of the molecular gas surface density in the bottom row, where shades of green depict the gas density in solar masses per square parsec. Together, these maps allow for direct comparison of where ionized and molecular gas coexist, separate, and evolve over time.

The H$\alpha$ emission is initially insignificant but rises quite quickly after a significant fraction of stars reach high stellar masses, roughly $5\,\Myr$ into the simulation (Figure~\ref{fig:fid_counting}); similar behavior is seen in the 24$\micron$ emission maps (Figure~\ref{fig:tf_ir}). 

Initially, the H$\alpha$ emission is relatively smooth and compact, akin to the classical Stromgren sphere \citep{stromgren_physical_1939}. In this phase, the emission is largely from dense cavities of ionized gas around massive stars. Beyond $8\,\Myr$, however, these cavities have expanded and the ionized gas density has dropped, suppressing the smooth component of H$\alpha$ (whose recombination emission goes as density squared). Instead, the H$\alpha$ emission becomes extended and highly structured. Its structure also becomes highly correlated with the remnants of the dense molecular gas. This correspondence suggests that at late times the high-surface density H$\alpha$ emission is primarily from ionization of dissociation fronts on the dense surfaces of the remnants of the original molecular reservoir, which by now has been disrupted or dispersed, leaving behind only a few regions that have survived the cumulative effects of massive star feedback.

Interestingly, the morphology of the H$\alpha$ emission at early times resembles the $24\mu$m mid-infrared emission shown in Figure~\ref{fig:tf_ir}. This similarity is not coincidental; both tracers are powered by radiation from the same population of embedded massive stars. The mid-IR emission arises from irradiated dust heated near the photoionization front, which are the same locations responsible for the strongest H$\alpha$ emission.

\begin{figure}
    \centering
    \includegraphics[width=0.48\textwidth]{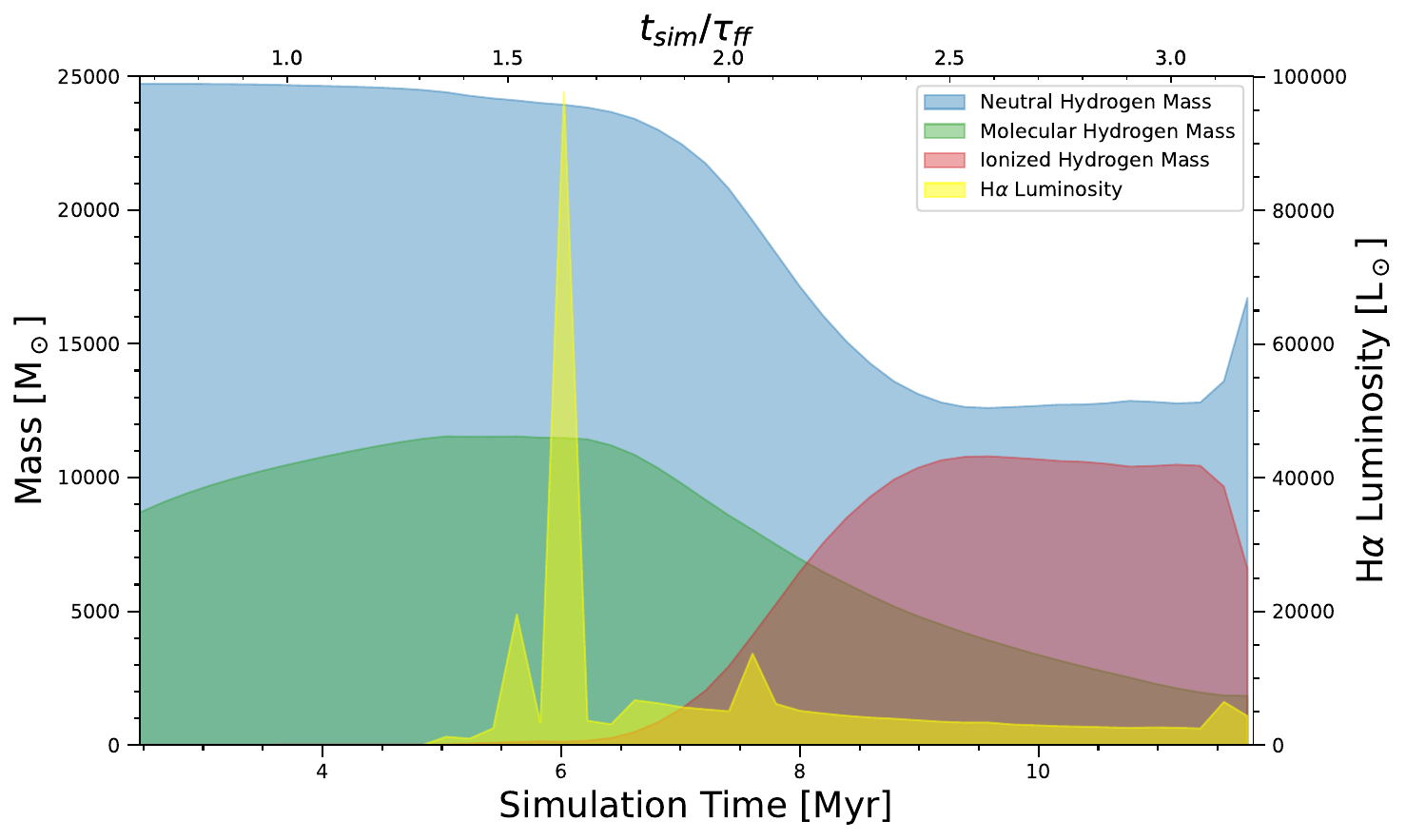}
    \caption{Time evolution of the gas mass in neutral (blue), and molecular hydrogen (green). The secondary axis shows the time evolution of the H$\alpha$ emission in L$_\odot$. }
    \label{fig:gas_mass_evolution}
\end{figure}

We also quantify the global evolution of the H$\alpha$ emission in Figure~\ref{fig:gas_mass_evolution}, in context with the various gas phases present in the simulation as a function of time. We plot the time evolution of the total H$\alpha$ luminosity (yellow curve, right axis), along with the total mass in molecular (green), neutral (blue), and ionized (red) hydrogen gas (left axis). Together, these curves display the interplay between gas phase transitions and emergent emission.

Confirming the visual impression from Figure~\ref{fig:h_alpha}, the H$\alpha$ luminosity remains negligible during the early stages of star formation (2.5–4.5 Myr), despite a growing population of low-mass protostars. The first strong rise occurs around 5 Myr, peaking sharply at 6 Myr, corresponding with the onset of significant massive stars. Following the brief peak, the luminosity declines more gradually, punctuated by brief secondary spikes at 7.7 Myr and 11.5 Myr. The former likely reflects a small burst of massive star formation, while the latter coincides with the first supernova event.

From 5 \Myr to 6 \Myr, the time-evolution of the H$\alpha$ emission fluctuates rapidly. While the full analysis for determining the exact cause of the large peak in the H$\alpha$ luminosity at 6 \Myr should be the focus of future studies, one possibility could be collisionally-ionized shocked regions from jets or winds. Another possibility is that the H$\alpha$ variability reflects short-timescale variability in the accretion of massive stars.

We also compare the evolution of the H$\alpha$ luminosity to the overall evolution of different gas phases in the simulation. Early on, the neutral gas mass steadily declines as material is converted into stars or accreted onto sink particles. Around 6 \Myr, the decline in neutral gas accelerates as feedback begins to ionize the cloud. This trend is mirrored by a rapid increase in the ionized gas mass, with the two phases clearly anti-correlated. In parallel, the molecular gas mass initially increases due to the collapse and cooling of the gas cloud, even as stars begin to accrete the dense gas. This growth continues until 6.2 \Myr, at which point the molecular reservoir begins to deplete after feedback from massive stars becomes significant. Some molecular gas does survive to late times, however, fueling a low level of residual star formation throughout the simulation.

We note that, surprisingly, the strongest H$\alpha$ emission actually precedes the formation of a significant ionized gas component in the cloud. As seen in Equation~\ref{eq:h_alpha}, recombination-driven H$\alpha$ depends on the square of the gas density, which is highest earlier in the simulation. At later times, there is more ionized gas, but most of that gas is now diffuse, with a much lower recombination rate, and thus a lower H$\alpha$ luminosity. This behavior echoes the picture developed in Section~\ref{sec:Drivers}, where H$\alpha$ is brightest when massive stars are still tightly coupled to localized gas, and fades once their feedback evacuates their immediate surroundings, even as global ionization of the cloud continues to grow.

\subsection{Optical and UV Emission from Stars} \label{sec:ob_uv}

As gas and dust are cleared by stellar feedback, the direct emission from stars begins to escape, allowing optical and ultraviolet light to emerge. These bands not only mark the transition from embedded to exposed phases of stellar formation, but also offer critical age constraints on the system. To evaluate how this starlight evolves and eventually becomes observable, we begin by computing the bolometric luminosity of each star, based on its mass and radius at each evolutionary stage (Section~\ref{sec:Drivers}).

\begin{figure}
    \centering
    \includegraphics[width=0.47\textwidth]{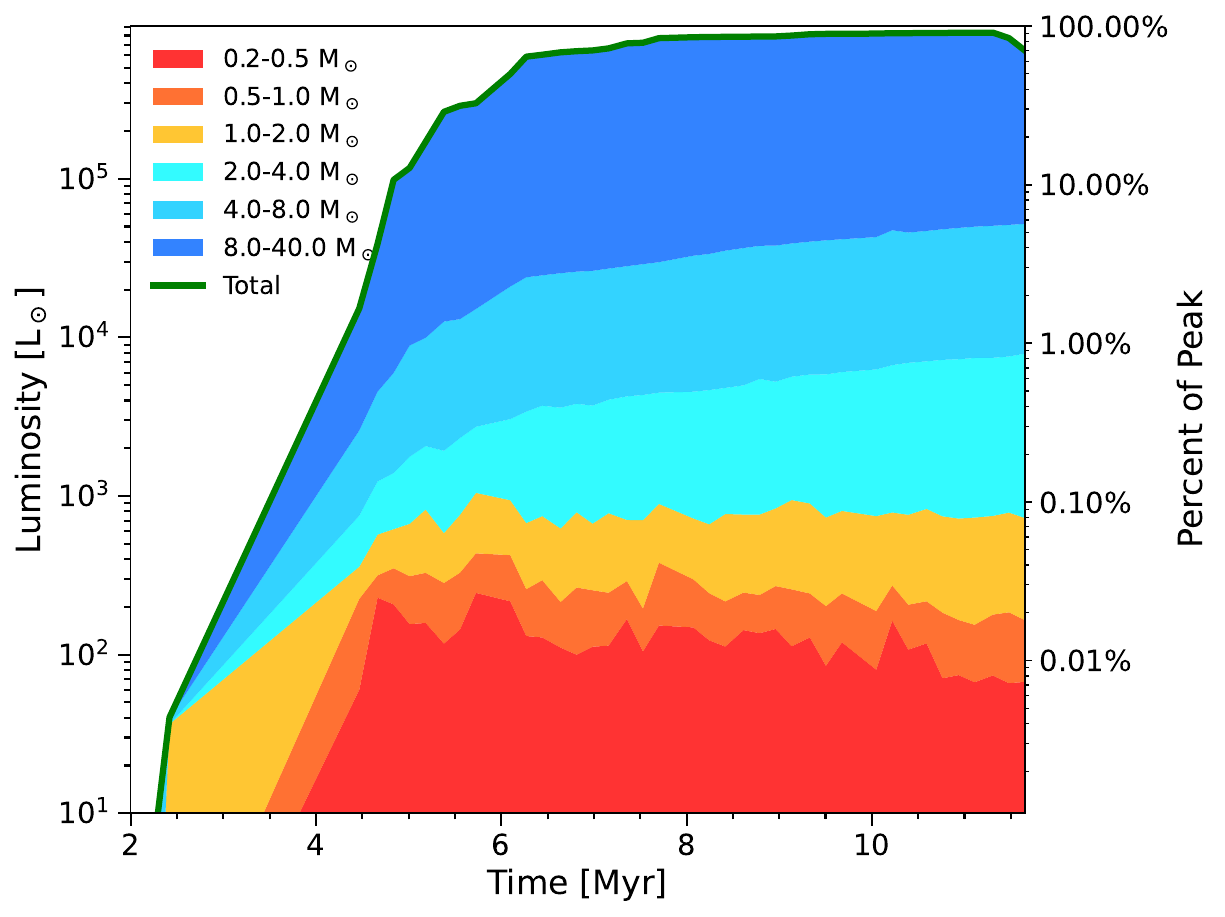}
    \caption{Bolometric luminosity of stars as a function of time. The green curve shows the total luminosity, while shaded regions indicate contributions from different stellar mass bins (same as in Figure~\ref{fig:stellar_extinction_tracks}). The right axis shows luminosity as a percentage of the peak total. Once they gain sufficient mass, around 4.2 \Myr (Figure~\ref{fig:fid_counting}), high-mass stars clearly dominate the luminosity the system.}
    \label{fig:sandpile}
\end{figure}

In Figure~\ref{fig:sandpile}, we decompose the total bolometric luminosity into contributions from stars in different mass bins. The green curve shows the total stellar output, while the stacked colored curves reflect how stars of different masses contribute over time (same bins and color scheme as Figure~\ref{fig:stellar_extinction_tracks}). These results reaffirm the key insight from Figure~\ref{fig:fid_counting}: once massive stars form—especially after $\sim$4.5 \Myr—they quickly dominate the cluster’s light budget, accounting for the majority of stellar luminosity well before the system is optically revealed.

While bolometric luminosity captures the full stellar energy output, what is detectable is shaped by dust attenuation and filter throughput. Therefore, we map the intrinsic stellar luminosities into synthetic observations of two filters frequently used to identify young stars: the \textit{GALEX} far-ultraviolet (FUV; 135–175 nm) and the \textit{HST} F555W optical band (458–620 nm). For each star, we integrate the stellar spectral energy distribution across the filter bandpasses.

To account for extinction, we determine the dust attenuation using the viewing angle-averaged ${\widetilde{A_{\rm V}}}$ values from Section~\ref{sec:measure_av}. Extinction in each band is then scaled according to the \citet{cardelli_relationship_1989} extinction law, with UV extinction taken to be $3.85\times{\widetilde{A_{\rm V}}}$. 

\subsection{The Relative Timing of Observational Indicators} \label{sec:dom_lum_timescales}

Having established how different star formation tracers emerge over the course of the simulation, we now turn to comparing their characteristic timescales. Each wavelength regime becomes visible on different timescales and persists for a different duration, depending on both the physical state of the cloud and the presence of massive stars. In this section, we compare when each tracer first appears, how long it dominates the association emission, and when it fades, providing a direct comparison between the observable windows offered by wavelength regimes.

To do this comparison, we consider quantities discussed throughout this section; dust luminosity (Section~\ref{sec:ir}), H$\alpha$ luminosity (Section~\ref{sec:h_alpha}), and the stellar luminosity as viewed in the optical and UV (Section~\ref{sec:ob_uv}. We present this comparison for potentially observed luminosities, where each quantity is measure through the specific telescope band pass with the corresponding wavelength attenuation.

\begin{figure*}
    \centering
    \includegraphics[width=0.99\textwidth]{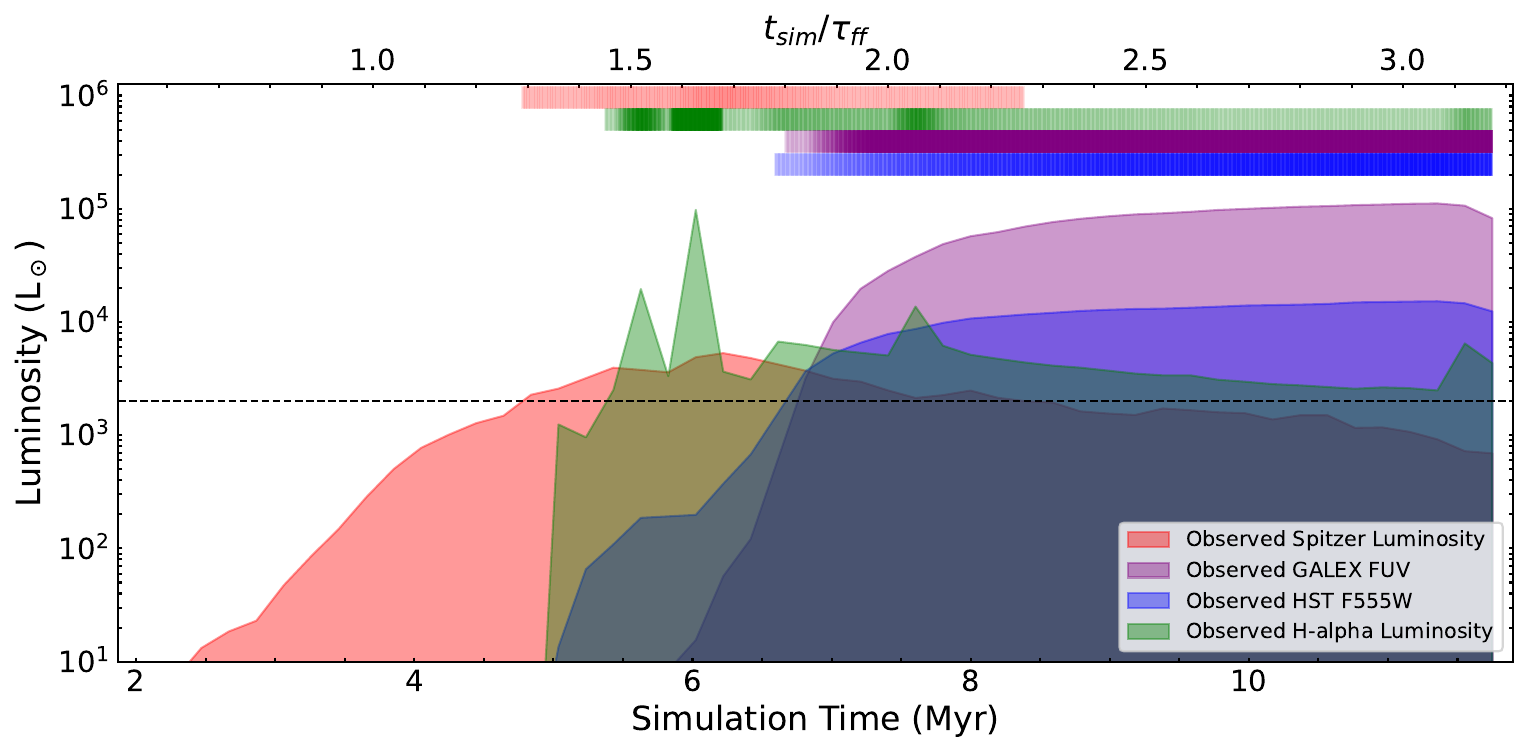}
    \caption{Observed luminosities in different telescope pass bands as determined by the median of 60 viewing angles. Red is 24 \micron\ emission observed in the \textit{Spitzer} band pass. Green is H$\alpha$ emission, while blue is optical light observed in the HST F555W filter, and purple is UV as observed by the GALEX FUV filter. The black dashed line at 2000 L$_\odot$ represents our threshold for ``onset'' timing, and the bars depict the times and relative brightness of each tracer with respect to this threshold. The mid-IR emission is most dominant for the first few \Myr, until the onset of massive star formation, when the H$\alpha$ emission dominates for $\sim1$\Myr. There is a brief period of $\sim0.5$\Myr where all tracers are present and bright, before the stellar luminosity becomes the dominant feature. }
    \label{fig:wavelength_comp}
\end{figure*}

We summarize the evolution of emission across wavelength regimes in Figure~\ref{fig:wavelength_comp}, which shows the observed luminosity from stars (blue and purple), dust (red), and ionized gas (green) as a function of time.  

Each tracer becomes prominent on a characteristic timescale, reflecting the physical conditions required for it to emerge. To make this comparison more concrete, Table~\ref{tab:tracer_times} lists the onset time, peak, and duration of each tracer’s dominance in the simulation. We define the onset to be when the luminosity crosses 2000 L$_\odot$. This relatively arbitrary threshold was chosen to best differentiate between when a tracer is present, and when it is clearly visible. While this threshold is unique to our current simulation, it adequately characterizes the intrinsic timescales for each tracer. 

\begin{table*}[ht]
\centering
\begin{tabular}{lcccc}
\hline
\textbf{Tracer} & \textbf{Onset (Myr)} & \textbf{Onset ($\tau_{\mathrm{ff}}$)} & \textbf{Duration (Myr)} & \textbf{Duration ($\tau_{\mathrm{ff}}$)} \\
\hline
\textit{Spitzer} 24 $\mu$m Dust Emission & 4.8 & 1.29 & 3.6 & 0.97 \\
656 nm H$\alpha$ Emission    & 5.4 & 1.45 & $>$6.4 & $>$1.72 \\
HST F555W Optical     & 6.6 & 1.78 & $>$5.2 & $>$1.39 \\
GALEX FUV    & 6.7 & 1.8 & $>$5.1 & $>$1.37 \\
\hline
\end{tabular}
\caption{Approximate emission timescales for each tracer, defined as the period above 2000 L$_\odot$. Times in $\tau_{\mathrm{ff}}$ use the fiducial $\tau_{\mathrm{ff}} = 3.7$ \Myr.}
\label{tab:tracer_times}
\end{table*}

The 24$\micron$ emission from irradiated dust appears first, describing the truly embedded phase of formation. It builds throughout and peaks around 6.2 \Myr, remaining prominently observable for nearly 3\Myr. H$\alpha$ emission emerges slightly later around 5 \Myr, and peaks rapidly, overlapping with the IR for about 1 \Myr before declining as gas is cleared. 

Although future work will be needed to determine the origin of this unexpectedly early peak in the H$\alpha$ luminosity, it winds up being nearly contemporaneous with the peak in the 24$\micron$ emission. This correspondence is surprising, given that the typical view is that the MIR/FIR traces a purely embedded phase that precedes any visible signatures of massive stars that later manifest in the H$\alpha$. Instead, both tracers have similar early peaks, and long tails. 

This unexpected behavior for the time-evolution of H$\alpha$ and 24$\micron$ emission again highlights the importance of local properties within the evolving cloud. Both luminosities will be strongly enhanced by having dense gas close to massive stars, although for different physical mechanisms -- i.e., recombination versus higher dust temperatures, to a degree that frequently dominates over the global evolution of the majority of the gas\footnote{Note the degree to which the high surface brightness 24$\micron$ emission in Figure~\ref{fig:tf_ir} is actually highly localized in small clumps, likely tracing associated young bright stars.}.

In contrast, optical and UV stellar light becomes dominant only after localized feedback has substantially cleared gas from the highest mass stars which dominate these wavelengths. This transition is rapid, occurring over just $\sim$0.5 \Myr. By 7.0 \Myr, over half of the intrinsic stellar light escapes without being absorbed, and direct starlight becomes the primary observable tracer.

We note that the specific values are subject to change with different decisions made throughout (i.e, threshold, dust-luminosity from secondary-emission), and therefore the derived timescale values should not be taken as ground truth. However, the overall evolution of these tracers are independent of these decisions, and the primary results of characteristic timescales are robust to the accuracy of the exact values. 

Together, these results reaffirm that dominant emission phases are not neatly sequential, but overlapping and are often controlled more by local feedback conditions than global cloud properties.

\section{Discussion}\label{sec:stats}
\subsection{Contextualizing STARFORGE and Observations} \label{sec:contextualizing}

Several previous works, both observational and theoretical, have attempted to quantify the timescales of the embedded stage of star formation. Compared to observation, one unique advantage of the work here is the plethora of information available in the STARFORGE simulations. Here we explore some notable differences between the two landscapes that should be kept in mind, discussing particularly relevant observations as examples, as well as ways that the results presented here can translate into the analysis of observations.

\subsubsection{The Timing and Sequence of Star Formation}

The model of star formation presented in the STARFORGE simulations indicates that stellar associations are not necessarily single age populations. We measure stars forming throughout all times of the simulations, with a standard deviation of 2.14\Myr and an rms of 6.7 \Myr Section~\ref{sec:age_determ}). Moreover, there is a clear mass progression in star formation, with low mass stars forming first, up to 1 Myr before high mass stars ($>$8\Myr) begin to form. We also note that some of the extended star formation is localized to a long-lasting and/or late-forming high-density gas region in the turbulent, but largely dissociated molecular gas cloud (Section~\ref{sec:location}).

However, most observational studies of embedded timescales typically assume stars form nearly instantaneously \citep[e.g.,][]{portegies_zwart_young_2010, hannon_h_2022, whitmore_phangs-jwst_2023, rodriguez_tracing_2025}. However, this assumption is increasingly challenged across a wide range of mass and spatial scales. In the most comprehensive studies of individual clusters like 30 Doradus \citep[e.g.,][]{de_marchi_star_2011}, or Orion \citep[e.g.,][]{bally_overview_2008, kounkel_apogee-2_2018}, often find evidence for extended star formation. Additionally, age spreads remain common when considering smaller associations and clusters  \citep[e.g.,][]{kerr_spyglass_2024}. Similarly, the extended star formation seen in STARFORGE echoes this broader observational trend, and is advocated in other theoretical works in both turbulence supported clouds \cite[e.g.,][]{krumholz_how_2020}, and collapse driven models \citep[e.g.,][]{vazquez-semadeni_turbulent_2024}. Together, these results suggest that the assumption of near instantaneous star formation may be an oversimplification in many systems.

While the single age stellar population is not an appropriate model for the behavior of the STARFORGE simulations as a whole, it is true that some of the extended star formation is spatially segregated from the earliest star formation (Section~\ref{sec:location}, \citet{guszejnov_effects_2022}). In an observational context, these sub-regions might be analyzed as independent regions, for which single age assumptions might be more appropriate approximations. However, in distant galaxies, these sorts of sub-regions are indistinguishable with current resolution, and cannot be analyzed independently. In such cases, it may be worth considering SED and spectral fits that allow a more extended or complex recent star formation history. Additionally, the simulations presented here use a limited set of initial conditions, and it may well be that the simultaneous, synchronous star formation assumption is more appropriate in other conditions.

We see some of these different approaches with respect to region size in the current literature. In some cases, there is sufficient resolution available to identify sub-regions that could behave like more traditional single age populations \citep{linden_goals-jwst_2024,knutas_feast_2025, rodriguez_tracing_2025}. In others, there is more of an attempt to characterize populations of entire star forming regions \citep[e.g.,][]{kerr_stars_2021,chevance_pre-supernova_2022, kim_environmental_2022, pan_gas-star_2022, kim_phangs-jwst_2023, ramambason_duration_2025}, which may contain a range of ages, more closely mimicking the behavior of the entire association presented here. These complementary approaches may well manifest in different inferred timescales for embeddedness, that can be brought into agreement by considering the different spatial scales and likely level of coherence in the analyzed regions.

Also pertaining to the sequence of star formation is the specific timing of the massive star formation, where quasi-instantaneous massive star formation is used in most timescale measurements \citep{schinnerer_molecular_2024}. However, the timing of massive star formation is currently an open question in the field \citep[e.g.,][]{kumar_youngest_2006, rosen_zooming_2020}. The massive stars forming after the smaller stars observed here has also been seen in other simulations \citep{vazquez-semadeni_multiscale_2024}. This timing supports the mode of formation where smaller cores form first, and high mass cores are a scaled up version of these smaller cores. 

\subsubsection{Interpreting Measures of Embeddedness for Individual Stars}

In analyzing the STARFORGE simulations, we identified two main pathways through which the embeddedness evolves. The first is the more traditional large scale gas clearance from the association (Section~\ref{sec:clst_bed}), driven by massive stars; within STARFORGE, this process has been analyzed in detail by \citet{farias_stellar_2024} across a range of cloud properties. 
The second is through highly localized feedback in the immediate neighborhood of individual massive stars (Section~\ref{sec:star_bed}). This latter process acts such that as embedded massive stars reach their maximum mass and luminosity (i.e., ending the accretion phase of their formation), their typical extinction drops dramatically, within less than 1\Myr. The emergence of low mass stars from the dust is much more gradual, more closely tracking the overall evolution of the cloud and their drift from regions of high gas density, rather than local feedback.

This behavior has some potentially interesting implications when analyzing embeddeness of individual stars. Typically, one measures spectra or spectral energy distributions, and uses stellar models to infer the line of sight extinction to the star. If massive stars are typically responsible for clearing out their own very local gas supply, they should rarely be seen with high extinctions outside of their initial accretion phase. Indeed, this has been seen in \citet{lindberg_dust_2024}, where populations of massive stars in M31 had a mean $A_{\rm V}$ of 1, independent of environment. We note, however, that the measures of embeddedness here are based on finding the median extinction over 4$\pi$ sterradians, which still allows individual stars to be highly embedded over up to 2$\pi$ of all viewing angles. As such, appropriate comparisons are with large enough samples that a range of viewing angles are represented.  

Another aspect that may differ when making observational assessments of stellar embedded timescales is the lifetime at the point of observation. In STARFORGE, we have complete access to the evolution of each individual star, including the entire proto-stellar phase (Figure~\ref{fig:gas_mass_evolution}). For massive stars, this phase terminates when gas accretion ends, which is also when the star becomes rapidly less embedded, synchronized through feedback that shuts off the local gas supply. As such, the lifetime of the protostar phase is roughly equivalent to the embedded timescale for massive stars, which is much shorter than the gas clearance timescale of the association as a whole. Therefore, the light from these stars will escape the system, and be visible at shorter wavelengths, even before gas clearance takes place.

\subsubsection{Interpreting Measures of Embeddedness for Semi/Un-resolved Association and Star Forming Regions}

In the absence of resolved stellar populations, clusters and associations are typically analyzed through their integrated spectral energy distributions. In STARFORGE, we find the expected sequence of emission that moves from luminous in the mid-IR, to a mixture of H$\alpha$ and mid-IR emission, to a brief window where optical stellar luminosity is comparable to the other tracers before the optical and UV light becomes dominant (Section~\ref{sec:dom_lum_timescales}. The timescales associated with these transitions are largely in agreement with those measured in external galaxies \citep[][]{kim_duration_2021,linden_goals-jwst_2024, rodriguez_tracing_2025,knutas_feast_2025, ramambason_duration_2025}. 

This agreement suggests that the same physical processes that govern light emergence in our simulations here are the same as those in the observed clusters. 
In particular, the simulations reveal that de-embedding is a highly local process, driven primarily by the feedback from massive stars disrupting their immediate surroundings, rather than the result of global cloud dispersal. Because such fine-grained spatial information is inaccessible in unresolved systems, the results presented here serve as a valuable calibration tool, offering physical insight into what integrated tracers are actually measuring. Observed timescales for light emergence can therefore be interpreted as reflecting the timing and locality of massive star formation, rather than a synchronized, global clearance of dust and gas.

For embedded clusters, there has been recent discussion about the embedded timescales of cluster formation being linked to cloud properties \citep[][\textit{Pedrini et al. in prep}]{dinnbier_how_2020, mcquaid_timescales_2024, knutas_feast_2025, ramambason_duration_2025}. Specifically, \citet{mcquaid_timescales_2024} and \citet{knutas_feast_2025} find the time a cluster stays embedded is inversely proportional to mass, where smaller systems stay embedded longer. 

We find that the overall timescales for formation and emergence scale strongly with the initial free fall time, such that the larger mass cloud has a larger characteristic timescales. We note however, that the simulations analyzed here only consider two masses with extremely similar initial conditions, such that the main difference is cloud mass and density. We also note that these timescales reflected the association as a whole, and that one might well derive different timescales when analyzing distinct sub-regions. In fact, when considering the embeddedness of individual stars in both clouds, we see the behavior of smaller stars staying embedded longer than massive stars (Section~\ref{sec:star_bed}). More instantaneous formation of stars would then manifest in the more massive complexes (i.e., higher number of massive stars) having shorter embedded timescales.

\subsection{Model Limitations} \label{sec:model_lim}

For numerical simulations to be computationally viable, certain trade off are made. A detailed discussion of the compromises in STARFORGE is available in \citet{grudic_starforge_2021}. In this section we discuss the specific ways in which this work is subjected to model limitations. 

First, the initial conditions within our STARFORGE model are an isolated, spherical cloud. Recent trends in the star formation literature tend to focus on the topic of triggered star formation, including cloud-to-cloud interaction, where gas is continually being deposited into the region where star formation is occurring \citep[e.g.,][]{fujita_high-mass_2021, fukui_cloud-cloud_2021, luisi_stellar_2021, wall_modeling_2020, zucker_compendium_2020}. This scenario would likely increase the gas removal time scale as more gas is being introduced to the system. Consequentially, the embedded timescales would likely be longer, allowing for more stars to form, and further increase the spread in stellar ages. 

Additionally, this interactive model of star formation has been argued as a potential formation mechanism of gravitationally bound star clusters \citep[e.g.,][]{xu_first_2020, wurster_gas_2023, sakre_massive_2023}. Because the model used in this work is of an isolated cloud, and does not form a gravitationally bound cluster, we cannot comment on the embedded timescales of such systems. 

Similarly, because the cloud in our model is isolated, we cannot comment on any potential galactic environmental dependencies on variations to embedded timescales. Recent work has suggested environment impacts the effectiveness of stellar feedback \citep[e.g.,][]{mcleod_impact_2021, della_bruna_stellar_2022, chevance_pre-supernova_2022, kim_environmental_2022}. Increased feedback would decrease the time stars stay embedded, and shorten the transition time while the inverse is true for decreased feedback. 

Similarly with regards to the galactic context, the simulations do not include a background galactic potential or vertical stratification, and therefore lack the additional confining weight and external pressure present in real galactic discs. In vertically stratified disc simulations, feedback-driven bubbles evolve in balance with the vertical weight of the ISM, and their ability to clear gas locally depends on this balance \citep[e.g.,][]{faucher-giguere_feedback-regulated_2013, ostriker_pressure-regulated_2022}. As such, stronger vertical gravity or higher ambient pressure can prolong the time that gas remains bound and dense, potentially lengthening the embedded phase. Our results should be viewed as characterizing the locally driven de-embedding physics, with the understanding that large-scale galactic context can modulate the absolute timescales.

We also dutifully acknowledge that the free fall time of the cloud is difficult to measure observationally, especially in dynamic environments. In these environments where one cloud can feed into another, the free fall time is an ever changing measurement, which may actually increase over time as mass is added to the system. Additionally, the free fall time is only a measure for the initial cloud. Once fragmentation occurs, each sub-clump has it's own free fall time, which changes with time as gas is consumed and expelled. While flawed, the free fall time presents the best mode the isolated cloud simulations in STARFORGE.

Despite these limitations, the work presented here is a reasonable model. STARFORGE presents one of the best star formation simulations due to its physical realism achievable with the high resolution. Importantly, it has been demonstrated to reproduce key phenomena such as star formation efficiency measurements and the IMF \citep{guszejnov_effects_2022, guszejnov_effects_2023}. While modeling star formation as an interactive process in conjunction with environmental surroundings is a realistic scenario, it also introduces an additional layer of complexity to the model, and there is still much to be learned about the formation process from extremely high resolution simulation of isolated clouds.


\section{Conclusion}
\label{sec:end}
In this work, we measure the mechanisms which drive the embbededness stages of star formation in the high resolution STARFORGE simulation. We look at the stellar extinction levels of both individual stars, and of the association as a whole. We also assess the time evolution of dust and H$\alpha$ emission, frequently used to identify embedded star formation. We find the following results:
\begin{itemize}
    
    \item The overall drop in extinction is driven by transitions experienced by individual stars. We show that stars all begin their lives with a high level of extinction, due to being embedded within their surrounding molecular cloud. The stars then experience a sharp, rapid decline in their typical extinction ${\widetilde{A_{\rm V}}}$ when they reach their maximum mass and stop accreting. Most notable is that this behavior persists no matter when a star forms. As a result, the overall evolution of association embedding is a reflection of this individual evolution, convolved with the history of star formation within the association.

    \item This transition in extinction is particularly sharp for massive stars, where the decrease to low extinction (${\widetilde{A_{\rm V}}}\!<\!1$) happens in less than half a free-fall time. For these massive stars, their maximum mass is reached simultaneously with their emergence from dust, as they continue to accrete up until their feedback overcomes the local accretion. Once these stars halt accretion, their localized, pre-supernova feedback then impacts the cloud as a whole, de-embedding lower mass stars in their vicinity. Therefore, the local feedback takes individual stars ${\widetilde{A_{\rm V}}}$ from 100 to 1, then the collective action of massive stars takes ${\widetilde{A_{\rm V}}}$ from one to zero as gas is cleared from the association as a whole. 

    \item Because of this effect the massive stars have on the association embeddedness, when massive stars form is particularly relevant for any measured timescale. We find the fraction of highly embedded stars begins to drop precipitously at the point where $\sim$50\% of the massive stars form, which equates to 1.6 $\tau_{ff}$ independent of cloud mass. After this drop in extinction begins, almost all stars are un-obscured within $\sim$3$\,\Myr$.
    
    \item We find a large age spread for when the stars formed in the simulation, with both low- and high-mass stars forming throughout the simulation, with a standard deviation of 2.14 \Myr, and rms of 6.7 \Myr. The star formation does have a well-defined peak, but there is also a long tail to late times, where star formation is sustained in dense sub-clumps that have survived earlier feedback.
    
    \item Overall, the fully embedded stage is brief, lasting only a few Myrs. While 24 \micron ~ emission is present before any other tracer, the peak in H$\alpha$ luminosity actually coincides with the peak of 24 \micron ~ luminosity. Transversely, the transition period where all tracers are visible and sufficiently bright occurs for 0.49 $\tau_{ff}$ (1.8 Myrs for fiducial cloud). Therefore overall, while IR wavelengths are an excellent tracer of star formation, the timescale where it is expected compared to other tracers is comparatively short. 

    \item Both 24\micron, and H$\alpha$ emission are brightest during the short interval when dense gas and dust remain tightly coupled to massive stars. At later times, dust and H$\alpha$ luminosities decline even as the ionized gas mass grows, enforcing the local clearing produced by high-mass stars as the dominant mechanism in setting embedded timescales. Additionally, the duration of the 24 \micron dominance agrees well with observationally inferred embedded timescales, reinforcing that the simulations correspond to the same physics that governs embedded tracers in real systems.
\end{itemize}

These results offer a new perspective on the evolution of embeddedness and the timescales at which stars remain enshrouded in their natal clouds, as viewed through stellar extinction, and multi-wavelength observation. Most notable deviations from previous studies include the age spread of stars, the effect localized feedback of massive stars has on embeddedness, and how the timing of when high mass stars form connects to the embedded timescale as a whole.

\begin{acknowledgements}
This work was completed as a part of the pre-doctoral fellowship at the Center for Computational Astrophysics. TW thanks the Flatiron Institute and the Simons Foundation. The Flatiron Institute is funded by the Simons Foundation. 
We graciously thank Angela Adamo, Cara Battersby, Lise Ramambason and Philip Myers for thoughtful conversations which helped shape this work. 
SSRO acknowledges support from NSF 2107340, NSF 2107942, a Peter O'Donnell Distinguished Researcher Fellowship, and a Donald Harrington Fellowship.
\end{acknowledgements}


\software{\texttt{astropy} \citep{astropy_collaboration_astropy_2013, astropy_collaboration_astropy_2018, astropy_collaboration_astropy_2022}, \texttt{Jupyter} \citep{kluyver2016jupyter}, \texttt{matplotlib} \citep{Hunter:2007}, \texttt{numpy} \citep{harris_array_2020}, \texttt{python} \citep{python}, \texttt{scipy} \citep{2020SciPy-NMeth, scipy_12522488}, \texttt{h5py} \citep{collette_python_hdf5_2014, h5py_7560547},
\texttt{skirt} \citep{baes_efficient_2011, camps_skirt_2015},\texttt{openai} \citep{openai2023gpt4}}

\bibliography{tobins_references, additional, software}

\appendix
\section{Supplementary Mock Image}\label{appendix}

\begin{figure*}[hb]
    \centering
    \includegraphics[width=0.98\textwidth]{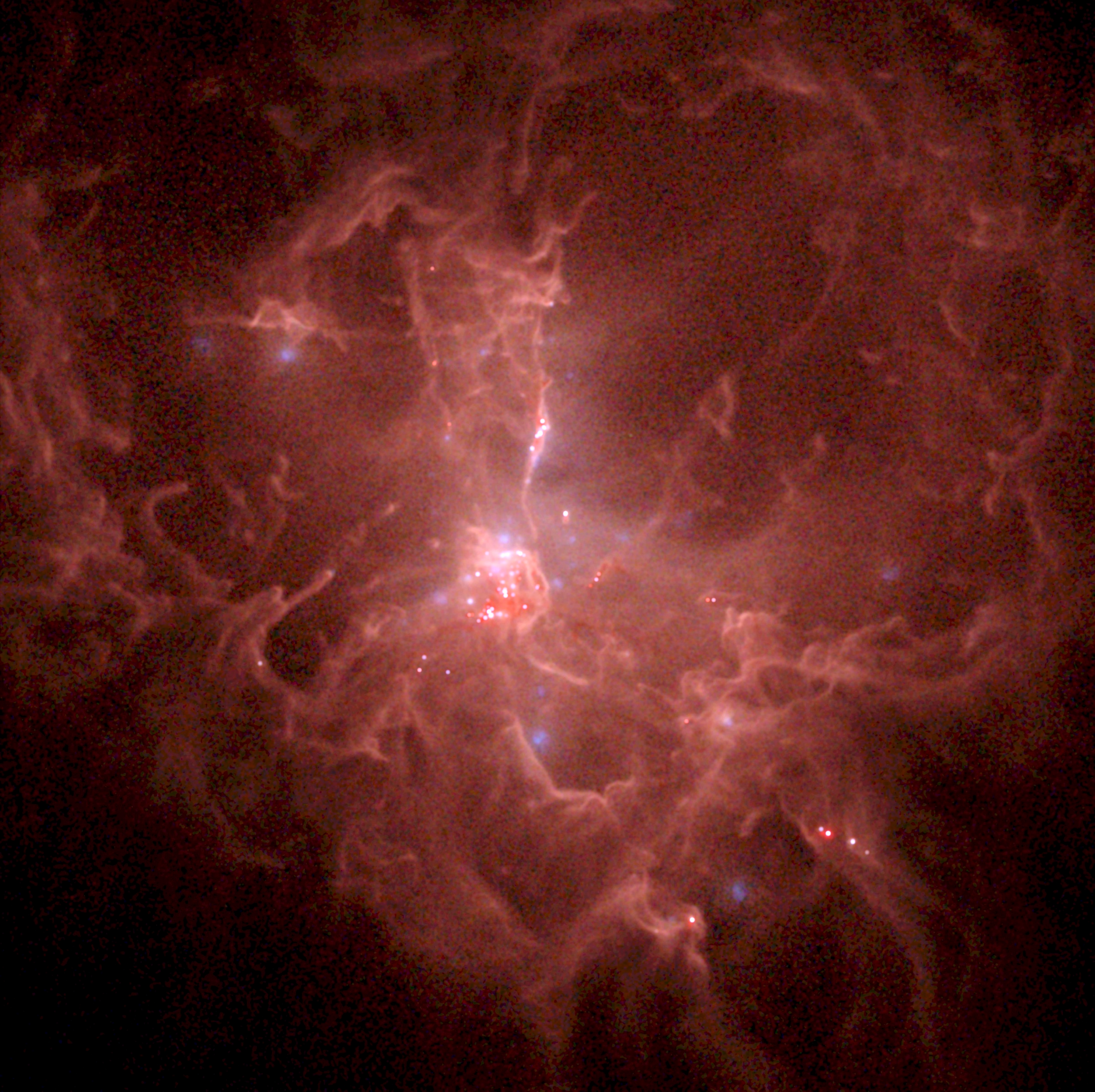}
    \caption{7.2 \Myr snapshot from the y-direction from Figure~\ref{fig:dust_lum_vs_time}, demonstrating the resolution available in the dust luminosity images. The full-high resolution version of this Figure can be found at \doi{10.5281/zenodo.17667459}}.
    \label{fig:apendix_snapshot}
\end{figure*}

We include a large, high resolution image of the dust luminosity from Figure~\ref{fig:dust_lum_vs_time} to highlight key features and provide the reader with the best visual representation of the data. Shown in Figure~\ref{fig:apendix_snapshot} is the snapshot at 7.2 \Myr, seen from the y-direction.

\end{document}